\documentclass[aip,reprint]{revtex4-1}

\usepackage[utf8]{inputenc} 
\usepackage[T1]{fontenc}    
\usepackage{hyperref}       
\usepackage{url}            
\usepackage{booktabs}       
\usepackage{amsfonts}       
\usepackage{nicefrac}       
\usepackage{microtype}      
\usepackage{graphicx}
\usepackage{lipsum}
\usepackage{comment}
\usepackage[caption=false]{subfig}
\usepackage{textcomp}
\usepackage{soul}           
\usepackage{amsmath}
\usepackage{listings}
\usepackage{chngcntr}
\usepackage{enumitem}
\usepackage[table,xcdraw]{xcolor}
\usepackage{booktabs}
\usepackage{multirow}
\usepackage{hhline}
\usepackage{amssymb}
\usepackage{siunitx}

\usepackage{euler}

\newcommand{\change}[1]{#1}

\definecolor{codegreen}{rgb}{0,0.6,0}
\definecolor{codegray}{rgb}{0.5,0.5,0.5}
\definecolor{codepurple}{rgb}{0.58,0,0.82}
\definecolor{backcolour}{rgb}{0.95,0.95,0.92}

\lstdefinestyle{mystyle}{
    backgroundcolor=\color{backcolour},   
    commentstyle=\color{codegreen},
    keywordstyle=\color{magenta},
    numberstyle=\tiny\color{codegray},
    stringstyle=\color{codepurple},
    basicstyle=\ttfamily\footnotesize,
    breakatwhitespace=false,         
    breaklines=true,                 
    captionpos=b,                    
    keepspaces=true,                 
    numbers=left,                    
    numbersep=5pt,                  
    showspaces=false,                
    showstringspaces=false,
    showtabs=false,                  
    tabsize=2
}
\let\oldtextbf=\textbf
\renewcommand*{\textbf}[1]{\ifmmode\mathbf{#1}\else\oldtextbf{#1}\fi}
\renewcommand*{\phi}[0]{\varphi}

\newcommand{\paragraphtitle}[1]{\textsf{\textbf{\small {#1}}}}

\draft 

\begin{document}

\title{Everything everywhere all at once: a probability-based enhanced sampling approach to rare events} 



\author{Enrico Trizio}
\affiliation{$^1$Atomistic Simulations, Italian Institute of Technology, 16156 Genova, Italy}


\author{Peilin Kang$^*$}
\email[]{peilin.kang@iit.it}
\affiliation{$^1$Atomistic Simulations, Italian Institute of Technology, 16156 Genova, Italy}

\author{Michele Parrinello$^*$}
\email[]{michele.parrinello@iit.it}
\affiliation{$^1$Atomistic Simulations, Italian Institute of Technology, 16156 Genova, Italy}


\date{\today}

\begin{abstract}
The problem of studying rare events is central to many areas of computer simulations. We recently proposed an approach to solving this problem that passes through the computation of the committor function, showing how it can be iteratively computed in a variational way while efficiently sampling the transition state ensemble. 
Here, we greatly ameliorate this procedure by combining it with a metadynamics-like enhanced sampling approach in which a logarithmic function of the committor is used as a collective variable. 
This procedure leads to an accurate sampling of the free energy surface in which transition states and metastable basins are studied with the same thoroughness. 
We show that our approach can be used in cases with the possibility of competing reactive paths and metastable intermediates. 
In addition, we demonstrate how physical insights can be obtained from the optimized committor model and the sampled data, thus providing a full characterization of the rare event under study. 
\end{abstract}


\maketitle 

\section{Introduction}
    Atomistic simulations hold the promise of allowing complex physicochemical processes to be studied with a minimum of theoretical assumptions.~\cite{frenkel2001understanding}   
    However, one of the major limits of this approach is the gap between affordable simulation time and the time scale of many highly relevant phenomena, such as chemical reactions, protein folding, or crystallization.  
    This is due to the presence of kinetic bottlenecks that slow down transitions between different states, which thus are called rare events. 
    The relevance of this problem is witnessed by the ever-growing literature on the subject.~\cite{Henin2022enhanced}
    
    In the last few decades, our group has contributed to this collective effort and very recently has proposed a new practical method to \change{ address} the rare event problem, which is based on computing the committor function $q(\textbf{x})$ and using it as a sampling tool, which we summarize in Section~\ref{sec:background}.~\cite{kang2024computing}  
    We recall here that the committor is a function of the atomic coordinates $\textbf{x}$ which, given two metastable states $A$ and $B$, is defined as the probability that a trajectory started at $\textbf{x}$ ends in $B$ without having first passed by $A$.~\cite{weinan2010transition}    
    The committor function is arguably the most precisely defined way of describing rare events and is a quantity that remains well-defined even if the transition from  $A$ to $B$  follows different competing pathways. As the committor has so many attractive features, much work has been done to determine it.~\cite{bolhuis2000reaction,ma2005automatic,berezhkovskii2005one, peters2006obtaining,e2006towards,rotskoff2022active,chen2023committor,chen2023discovering,li2019computing,jung2023machine,mitchell2024committor,he2022committor,khoo2019solving}

    To evaluate $q(\textbf{x})$  we have taken advantage of the variational principle of Kolmogorov,~\cite{kolmogoroff1931analytischen} which states that if the boundary conditions $q({\textbf{x}}_A)=0$ and $q({\textbf{x}}_B)=1$ are satisfied for configurations $\textbf{x}_A$ and $\textbf{x}_B$ from the metastable basins, the committor can be computed by minimizing the functional:
        \begin{equation}
          K[q(\textbf{x})] = \langle |\nabla_\textbf {u} q(\textbf x)|^2 \rangle _{U(\textbf{x})}  
        \end{equation}
    where the average $\langle \cdot \rangle_{U(\textbf{x})}$ is over the Boltzmann distribution associated with the interatomic potential $U(\textbf{x})$ at inverse temperature $\beta$, and $\nabla_\textbf{u}$ denotes the gradient with respect to the mass-weighted coordinates.
    
\begin{figure*}[t!]
        \begin{minipage}{\linewidth}
            \centering
            \includegraphics[width=0.9\linewidth]{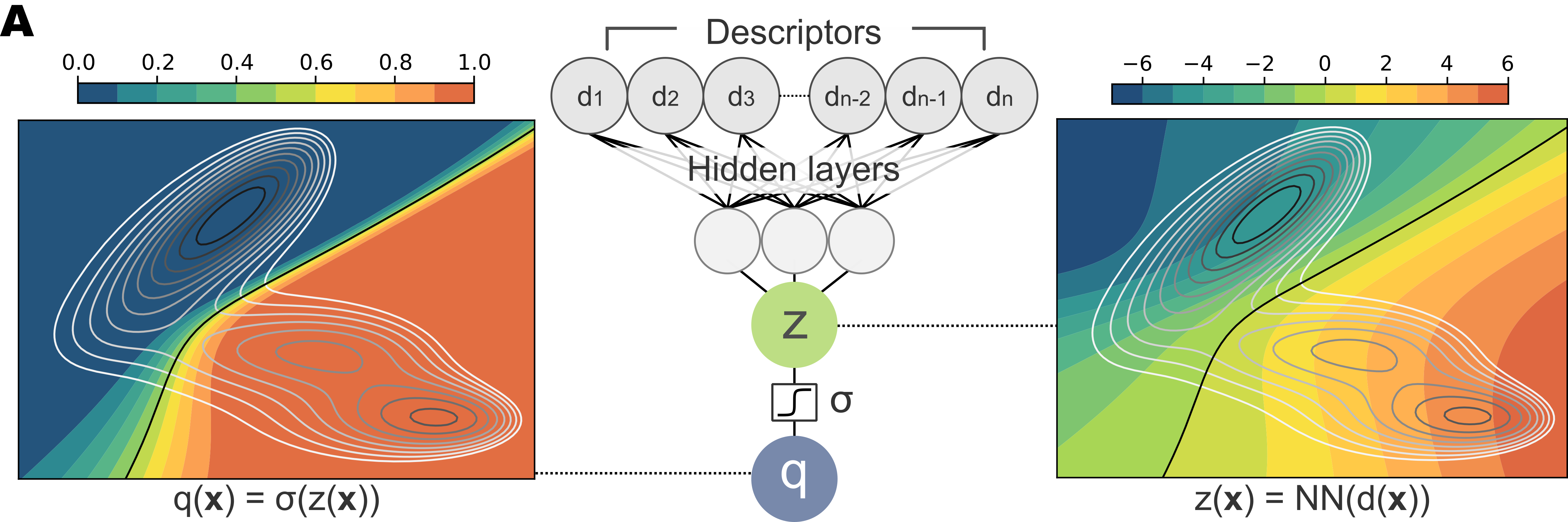}
        \end{minipage}
        \vfill{\vspace{0.4cm}}
        \begin{minipage}{\linewidth}
            \centering
            \includegraphics[width=0.9\linewidth]{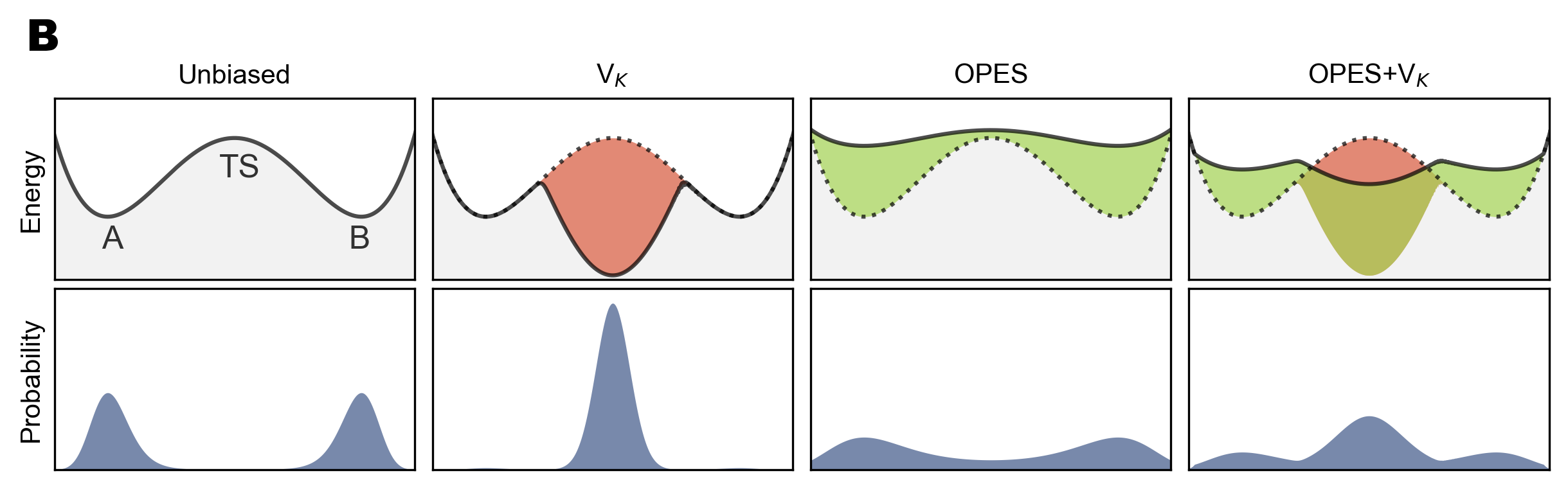}
        \end{minipage}
        \caption{\textbf{Schematic representation of enhanced sampling framework.} \textbf{A}. Relationship between the committor function $q$ and the $z$ CV. 
        In the model, a set of input descriptors $\textbf{d}(\textbf{x})$ is combined using a neural network into a one-dimensional latent space $z$.
        The final output that, upon optimization, represents the committor function is obtained by applying a sigmoid-like activation function $\sigma$ on $z$.
        As a consequence, $z$ and $q$ encode the same information, but the $z$ space is more spread and smooth with respect to the sharper $q$ space, and it is thus more suitable to be used in an enhanced sampling context.
        \textbf{B}. Different scenarios from the energy and probability points of view (top and bottom rows, respectively) for a double-well model potential with two states, A and B.
        Left to right, we show the unbiased scenario, a biased scenario where the metadynamics-like OPES bias fills the basins to promote transition, a biased scenario where the Kolmogorov bias ($V_K$) stabilizes the transition state (TS), a biased scenario in which the combined bias OPES+$V_K$ allows extensively sampling the metastable and transition states and promoting transitions between them.}
        \label{fig:NN_z_q}
    \end{figure*}

    In Ref.~\citenum{kang2024computing}, we computed the variational minimum of $K[q(\textbf{x})]$ using a self-consistent iterative procedure that to be started, only requires that the initial and final states are known.  
    Key to the success of this approach has been the use of a bias potential $V_K$ that is a functional of the committor
        \begin{equation}
            V_K(\textbf{x})=- \frac{1}{\beta} \log (\vert \nabla q(\textbf{x})\vert^2)
        \end{equation}
    which is able to focus sampling on the otherwise elusive transition state region (see Figure~\ref{fig:NN_z_q}).  
    This approach has proven to be highly successful in finding and extensively sampling the transition state ensemble (TSE). 
    However, additional simulations were needed to evaluate the system free energy since $V_K(\textbf{x})$ disproportionately favors sampling of the transition state ensemble region.
    This additional calculation can be performed, for instance, using the data collected during the iterative process to generate an efficient collective variable~\cite{bonati2021deep, ray2023deep} (CV) and then performing a metadynamics-like calculation using this CV.~\cite{laio2002escaping, invernizzi2020rethinking, trizio2024advanced}
    
    However, in principle, such a post-processing search for a good CV should not be necessary since the committor has been argued to be the best possible one-dimensional reaction coordinate.~\cite{berezhkovskii2005one, peters2006obtaining,ma2005automatic,li2014recent,he2022committor}
    Unfortunately, directly using $q(\textbf{x})$ as a CV is unpractical, if not impossible, in an enhanced sampling context.
    Indeed, in the basins, where $q(\textbf{x})$ is either $\approx 0$ ($\textbf{x} \in A$) or $\approx 1$ ($\textbf{x} \in B$), different configurations are distinguished only by very tiny numerical variations of $q(\textbf{x})$, thus leading to numerical difficulties.
    Furthermore, another source of numerical issues comes from the very sharp behavior of $q(\textbf{x})$ in the transition state region,~\cite{khoo2019solving} further complicating its use as a CV (see Figure~\ref{fig:NN_z_q}A and Supplementary Figure~\ref{sup_fig:numerical_issues}). 

    However, a convenient solution can be found if we recall that in Ref.~\citenum{kang2024computing}  $q(\textbf{x})$ was expressed as a neural network whose output $z(\textbf{x})$ was fed to a step-like activation function $\sigma$ to facilitate imposing the right functional form (see Figure~\ref{fig:NN_z_q}A)
        \begin{equation}
            q(\textbf{x}) =  \sigma \big ( z(\textbf{x}) \big)
        \end{equation}
    The output of the network $z(\textbf{x})$ encodes the same information as $q(\textbf{x})$ but is smoothly varying and can be used in practice as an efficient CV.
    This observation is crucial to the success of the improved approach we present here, which makes the method of Ref.~\citenum{kang2024computing} an effective and semi-automatic way to perform free energy calculations. 
    Indeed, we do not wait for the end of the iterative process to perform an additional free energy calculation, but, \change{during the whole iterative procedure}, we \change{simultaneously apply the} $V_K$ \change{potential and} an On-the-fly Probability Enhanced Sampling~\cite{invernizzi2020rethinking, trizio2024advanced} (OPES) bias that uses $z(\textbf{x})$ as a CV \change{that} it's aimed at filling the energy landscape to promote transitions (see Figure~\ref{fig:NN_z_q}B).  
    We shall refer to this combined bias as OPES+$V_{K}$, and we show that this, alongside a few technical improvements, accelerates the convergence of the iterative procedure and leads to a balanced sampling, in which configurations belonging to the metastable states and to the transition state ensemble are equally well sampled, and no additional calculations are needed to obtain a converged free energy estimate \change{upon reweighting of the sampled configurations by the total bias} (see Figure~\ref{fig:NN_z_q}B).

    For illustrative purposes, we first demonstrate the workings of our approach on some simple but informative systems like some two-dimensional models and the classical example of alanine dipeptide in a vacuum.
    We then move to more complex systems, studying the folding of the chignolin protein in water and the interaction between an organic ligand and a calyxarene, which is a simplified but non-trivial model for drug-protein interaction. 
    In this last example, different binding pathways are possible, and the ligand visits several intermediate states on its way to the binding pocket.
    This demonstrates the generality of our approach, which does not require assuming that there is only one reaction pathway.
    A full description of the methods, a summary of the procedure, and other technical details are provided in Section~\ref{sec:methods} and in the Supporting Information.

\section{Results}

\subsection{M\"uller potential}

    \begin{figure}[t!]
        \centering
           \includegraphics[width=0.9\linewidth]{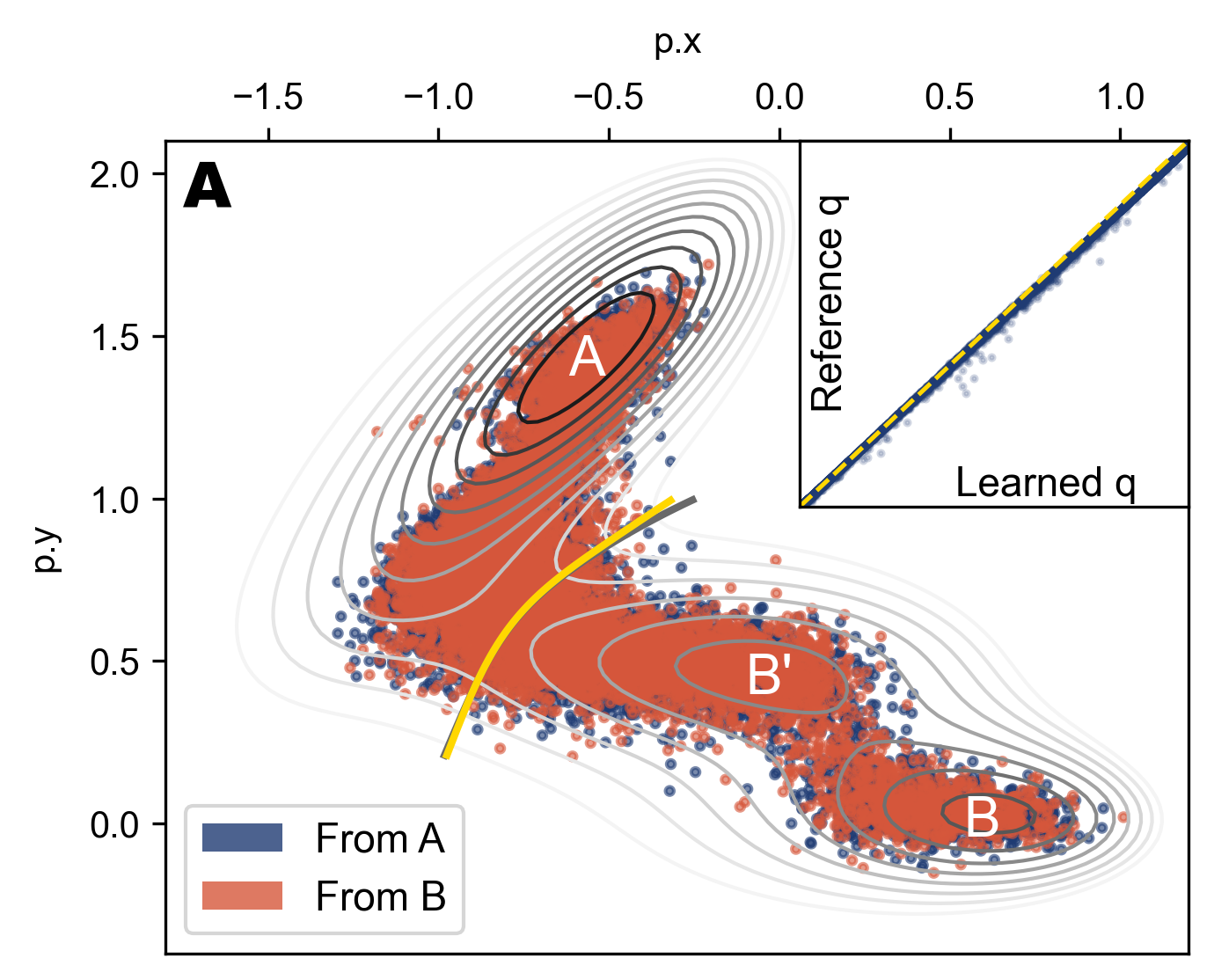}
           \includegraphics[width=0.9\linewidth]{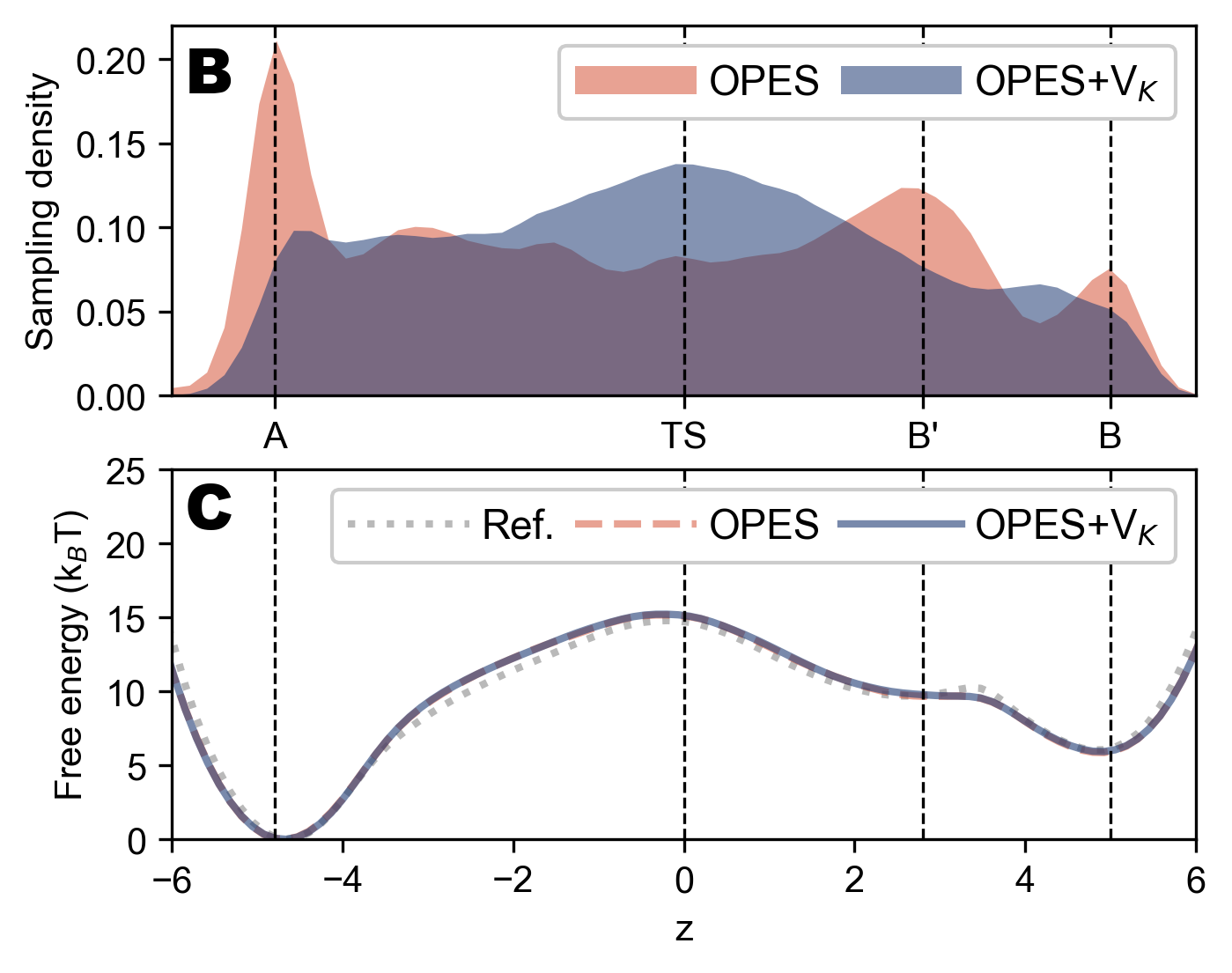} 
        \caption{\textbf{M\"uller-Brown potential}. 
        \textbf{A}. Scatter plot of the sampling under the coupled action of the OPES bias along the $z$ CV and the TS-oriented Kolmogorov bias $V_K$ at convergence after two iterations of our protocol.
        The isolines outline the energy levels of the potential, and points sampled starting the simulation from basins A and B are reported in blue and red, respectively.
        To demonstrate convergence, we report the physically relevant part of the $q\sim0.5$ isoline for a numerical reference for the committor (grey line), and the one learned from our model (yellow).
        The outputs of the same two functions on all the sampled points are compared in the inset, in which the yellow dashed line gives the equivalence line.
        \textbf{B}. Distribution of the sampled configurations along the $z$ CV for simulations driven by OPES+$V_K$ (blue curve) and only by OPES (red curve). 
        The black dashed lines indicate the relative peaks of the two distributions.
        \textbf{C}. Free energy estimate along $z$ obtained from the two simulations reported in \textbf{B} and from a numerical reference (grey dotted line). The black dashed lines are consistent with those reported in \textbf{B}.
        }
        \label{fig:muller}
    \end{figure}

    We start discussing the merits of our new approach on the simple but instructive example of a single particle diffusion on the M\"uller-Brown potential.
    As in our previous work,~\cite{kang2024computing} we use the $x$ and $y$ coordinates of the particle as descriptors, thus allowing a fair comparison between the two methods.
  
    Performing OPES+$V_K$ simulations allows covering the whole relevant phase space from the metastable states through the TS, as shown in Figure~\ref{fig:muller}A. 
    Moreover, to achieve this better sampling, it is not necessary to wait until a converged model is available, as good sampling is obtained even from the very first iteration (see Supplementary Figure~\ref{sup_fig:muller_sampling}) in which the model is not much more than a classifier trained to satisfy the boundary conditions $q({\textbf{x}}_A)=0$ and  $q({\textbf{x}}_B)=1$ (see Section~\ref{sec:learning_committor}).
    However, this result should \change{not surprise}, as machine-learning classifier-like CVs have already proved effective in driving enhanced sampling simulations with OPES.~\cite{bonati2020data, trizio2021enhanced, ray2023deep}
    In contrast, using only $V_K$ in Ref.~\citenum{kang2024computing}, the sampling was unbalanced toward the TS region and often heavily dependent on the basin from which the simulation was started.
    Another advantage in terms of sampling comes from the action of the Kolmogorov bias $V_{K}$, which favors extensive sampling of the TS region. 
    This is evident if one compares the distribution along $z$ of the points sampled with a standard OPES simulation and one performed by combining OPES and $V_K$ (see  Figure~\ref{fig:muller}B).
    From the two distributions, it can be clearly appreciated the different emphasis that the two simulations put on the TS region.
    When only OPES is used, most of the time is spent in the metastable basins. On the other hand, with the combined bias, a large number of TSE configurations is harvested.
    Such configurations are precious not only for better investigating the TS properties but also for effectively applying the variational principle to optimize the committor.

    \begin{figure*}[t!]
        \centering
        \includegraphics[width=1
        \linewidth]{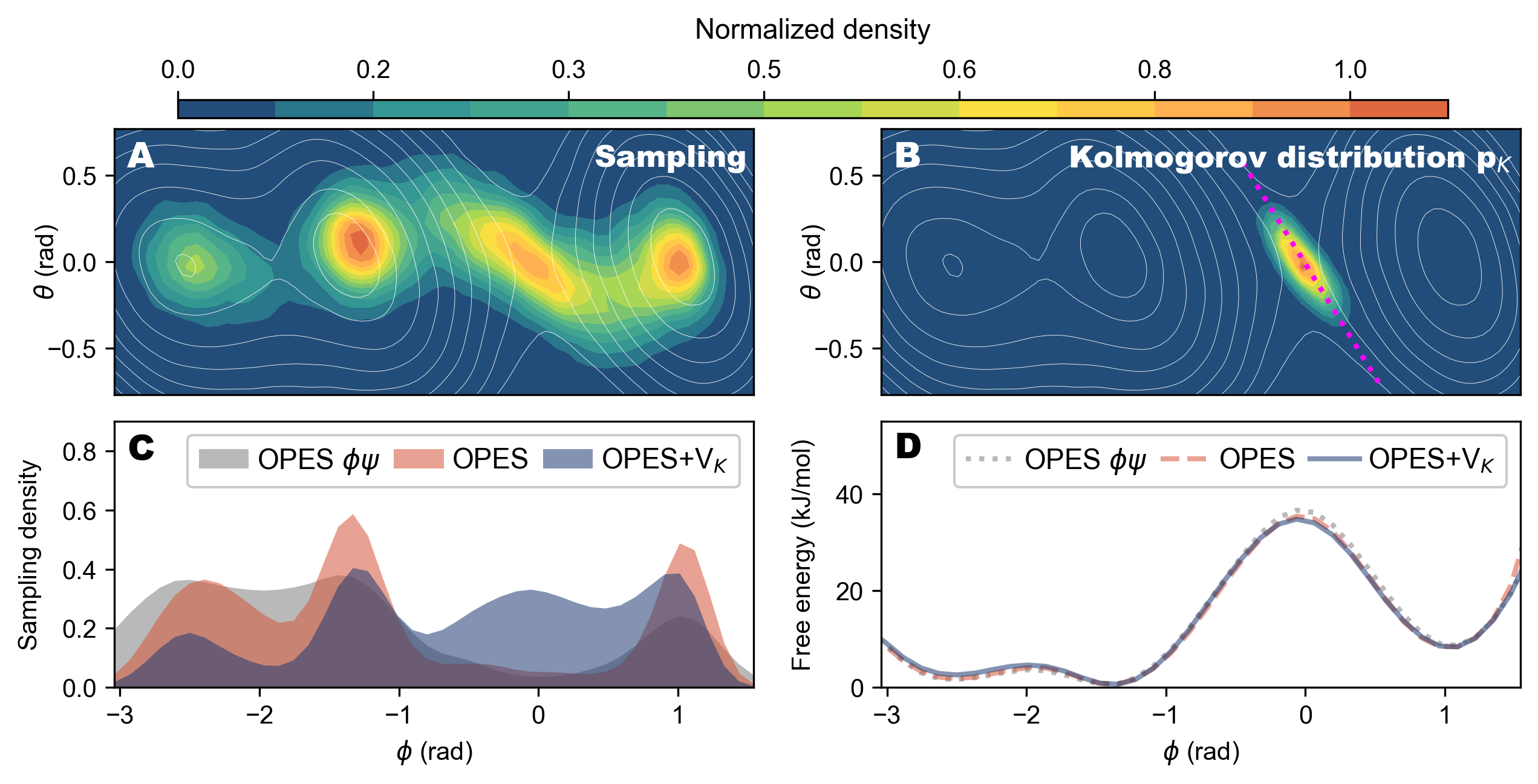}
        \caption{\textbf{Alanine dipeptide}.
        Top row: normalized densities in the plane of $\phi$ and $\theta$ torsional angles of the sampled configurations (\textbf{A}) under the joint action of the OPES bias on the $z$ CV and the TS-oriented Kolmogorov bias $V_K$ and of the corresponding reweighted Kolmogorov distribution $p_K$ (\textbf{B}). 
        The white isolines provide the reference free energy levels, whereas the magenta dotted line in \textbf{C} highlights the relation between $\phi$ and $\theta$ for TSE configurations proposed in Ref.~\citenum{bolhuis2000reaction}.
        Bottom row: comparison of the distribution of sampled configurations (\textbf{C}) and free energy estimates (\textbf{D}) along $\phi$ with different enhanced sampling setups: OPES on $\phi$ and $\psi$ (grey), OPES on $z$ (red), and OPES+$V_K$ on $z$ (blue, same data as \textbf{A}).}
        \label{fig:alanine}
    \end{figure*}
    
    In addition, a second advantage of this approach is related, besides the \textit{quantity} of the sampling, to its \textit{quality}.
    Thanks to the static nature of $V_K$ and the robust OPES framework,~\cite{invernizzi2022exploration, trizio2024advanced} which is designed to quickly converge to a quasi-static regime, very reliable weights can be assigned to the sampled points, thus allowing for a proper reweighting to the true Boltzmann distribution, which, again, has several implications.
    On the one side, this allows the recovery of the free energy surface (FES), as we show in Figure~\ref{fig:muller}C.
    There, we report the free energy profiles along $z$ obtained by reweighting the same two trajectories considered above, compared with an analytical reference.
    The three curves are almost indistinguishable, thus proving the quality of the reweighting procedure and the (non-obvious) absence of detrimental effects of the $V_K$ bias on such free energy calculations.

    On the other side, having such reliable weights also for the points in the metastable regions facilitates the convergence of the iterative procedure, at variance with Ref.~\citenum{kang2024computing}.
    Indeed, even for this simple example, this approach allows converging the estimate of the committor to the numerical reference almost perfectly in just two iterations (see inset of Figure~\ref{fig:muller}A).

    \subsection{Alanine Dipeptide}

    In most publications devoted to rare event sampling, it is traditional to illustrate the capability of a new method on the conformational equilibrium of alanine dipeptide, and here, we shall stick to this tradition.
    Moreover, having studied alanine also in our previous work,~\cite{kang2024computing} we are in the position to assess the improvement in convergence speed due to our improved protocol.
    There, a good guess for the committor function was obtained from the $6^{\text{th}}$ iteration of the procedure, whereas here, it was already available with half of the iterations (see Supplementary Table~\ref{sup_tab:alanine_iterations}).
    This allowed us not only to efficiently sample the whole relevant phase space (see Figure~\ref{fig:alanine}A) but also to properly identify the TSE (Figure~\ref{fig:alanine}B), which is known to exhibit a linear relation between $\phi$ and $\theta$.
    ~\cite{bolhuis2000reaction}   
    It is also worth noting that one can still estimate the Kolmogorov distribution $p_K$ by reweighting the OPES+$ V_K(\textbf{x})$ sampled configurations only with the OPES part of the bias. 
    As discussed in Ref.\citenum{kang2024computing}, this allows defining in a precise way the TSE. 

    Also in this case, it is instructive to compare the distribution of points sampled using different bias potentials. 
    In Figure~\ref{fig:alanine}C, we compare the outcome of three different OPES simulations, in one, we use the traditional $\phi$ and $\psi$ as CVs, in the second the  $z(\textbf{x})$ is used as CV, and finally in the third the action of the $z(\textbf{x})$ OPES bias is boosted by the Kolmogorov bias $V_K(\textbf{x})$. 
    As seen for the M\"uller potential, combining the $V_K$ potential with OPES systematically enhances the sampling of the TSE while still guaranteeing a thorough sampling of the metastable states.
    From such data, we can thus \change{correctly estimate} the free energy of the system, which is indistinguishable not only from that obtained in the normal OPES run on $z$ but also from the reference simulation using $\phi$ and $\psi$ (see Figure~\ref{fig:alanine}D), thus increasing our confidence in the method.

    \subsection{Double Path Potential}
    \label{sec:double_path}
    \begin{figure}[h!]
        \centering
        \includegraphics[width=0.9\linewidth]{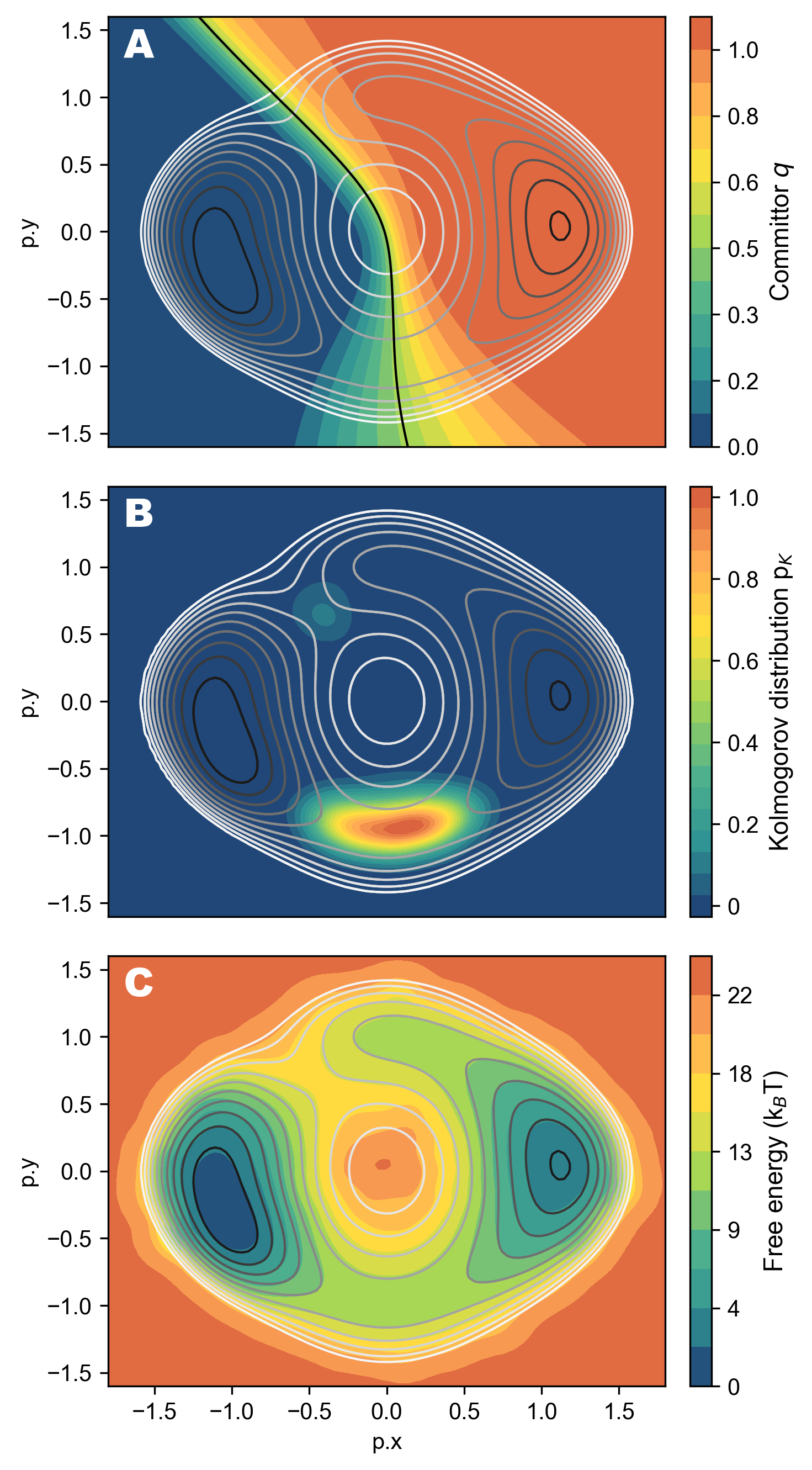}
        \caption{\textbf{Asymmetric double path potential}.
        \textbf{A}. Contour plot of the learned committor function superimposed to the isolines of the potential.
        \textbf{B}. Kolmogorov distribution $p_K$ computed from a simple reweighting of the sampled trajectory by the OPES bias only.
        \textbf{C}. Two-dimensional free energy surface recovered by reweighting the sampled trajectories by the total effective bias OPES+$V_K$.}
        \label{fig:double_path}
    \end{figure}
    
     Another situation that can occur in practice and which is not straightforward to deal with using other methodologies is the one in which the reaction can follow different pathways~\cite{vlugt2001pathways, borrero2016avoiding, bolhuis2018nested, mandelli2020paths, capelli2019exhaustive}. 
     A simplified and yet instructive model is the one depicted in Figure~\ref{fig:double_path}, where the system can go from state $A$ to state $B$ following two different pathways, with the upper path characterized by a larger free energy barrier. 
     
     With our approach, both paths are sampled starting from the first iteration, thus allowing for a quick convergence of the procedure (see Supplementary Figure~\ref{sup_fig:double_path_sampling}).
     It is then instructive to focus on the final results reported in Figure~\ref{fig:double_path}.
     In panel A, we report the converged committor function for the system, showing not only that both paths can be handled at the same time but also that the committor varies more abruptly in the case of the higher barrier, whereas it is more spread for the favored channel.
     This behavior provides a non-trivial example of why our choice to use the Kolmogorov distribution~\cite{kang2024computing} $p_K$ (see Eq.~\ref{eq:kolmogorov_distribution}) to identify TSE is to be preferred over the conventional $q\simeq0.5$ definition.
     If one considers the $p_K$ for this system (see Figure~\ref{fig:double_path}), one can readily see that most of the contribution to the transition flux comes from the lower channel, and only a minor portion goes through the other. 
     In contrast, the $q\simeq0.5$ criterion would have given the same importance to the two paths and also to much higher energy regions that were actually never sampled, not even in biased simulations (see Supplementary Figure~\ref{sup_fig:double_path_sampling}).
     The predominance of the lower path is also reflected by the distribution of the sampled points, which are much denser there (see Supplementary Figure~\ref{sup_fig:double_path_sampling}).
     \change{Eventually, in more complex cases, a more homogeneous sampling could also be achieved by biasing additional orthogonal CVs that could distinguish them, which is also straightforward in the OPES framework (see Section\ref{sec:committor_cv}).}
     Nevertheless, \change{in this case,} after reweighting, an almost perfect 2D FES for the system, including precise estimates of the barriers, can be recovered effortlessly, as shown in Figure~\ref{fig:double_path}C.

     It is worth noticing that all these results were obtained without any prior information on the system if not knowledge of the initial and final metastable states.
     From this little amount of information, through the iterative procedure and thanks to the combined action of the OPES+$V_K$, we could not only blindly discover the two possible paths but also fully characterize them.
     Another more realistic example of this situation will also be shown in the following.
     
\begin{figure*}[t!]
    \centering
    \includegraphics[width=1\linewidth]{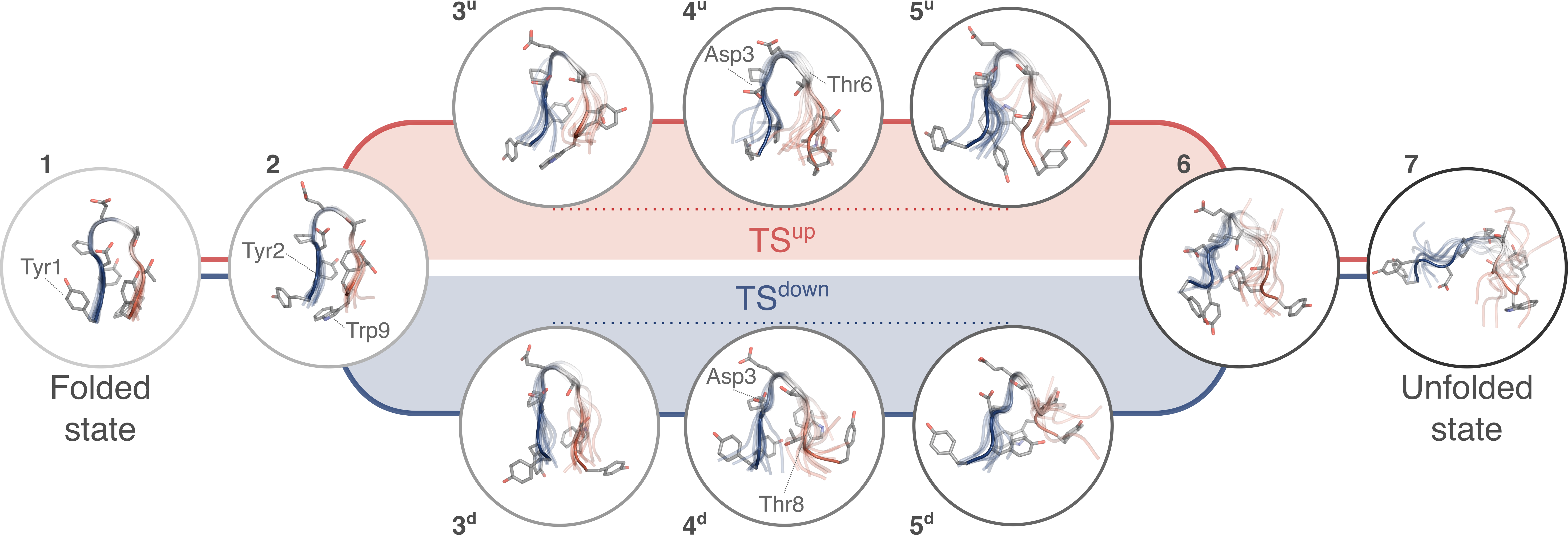}
    \caption{\textbf{Chignolin folding.} Snapshots of representative configurations along the (un)folding pathway of chignolin protein binned according to the $z$ CV value.
    Each frame reports in solid color the medoid configuration of the corresponding bin and, in transparency, the superimposition of 20 random configurations from the same bin to show the internal variance. 
    Carbon atoms are depicted in gray, oxygen in red, nitrogen in blue, and hydrogen in white.
    The main chain of the protein is depicted as a cartoon colored from blue to red according to the residue number from 1 to 10, whereas the side chains are reported in transparency in licorice for the medoid configuration only.
    For the TS region, two routes are highlighted based on clustering analyses, one in which only the upper part of the H-bond network of the folded structure is formed (TS$^{\text{up}}$, red path), in the other, only the lower part is formed (TS$^{\text{down}}$, blue path).
    Relevant residues along the transitions are indicated with grey labels.}
    \label{fig:chignolin}
\end{figure*}

\subsection{Chignolin: Protein Folding}
    A more challenging test for our method is the study of the folding of the solvated chignolin protein into a stable hairpin structure.
    Conveniently, for this system, results from long unbiased MD simulations are also available,~\cite{lindorff2011fast} providing a reference for the folding energy and the overall dynamics of the process.
    In our previous work,~\cite{kang2024computing} we already studied this system, characterizing the TSE committor and identifying two possible transition routes.
    There, we used as descriptors the 45 distances between the alpha carbon atoms, which were able to describe well the protein backbone.
    Here, we make use of the greater efficiency of our approach to enlarge the number of descriptors, using the 210 distances selected in Ref.~\citenum{bonati2021deep} that also include the interaction of the backbone with the side chains.      
    As discussed in Ref.~\citenum{kang2024computing}, better variational flexibility typically leads to more accurate results and to a lower value of the variational bound $K_m$.
    Here, in fact, on the same dataset, the minimum goes from the $K_{m}\approx 19$ to the much lower bound of $K_{m}\approx 2.2$ (see also Supplementary Section~\ref{sup_tab:chignolin_iterations}).
    
    Such a remarkable quantitative difference confirms that adding side-chain information greatly improves the description of the process.   
    This is testified by the fact that we can obtain an accurate estimate of the folding energy with a set of 1 $\mu$s-long simulations in the last iteration of the procedure, which is in good agreement (that is, less than 1$k_BT$) with the results from the 106 µs long unbiased simulation of Ref.~\citenum{lindorff2011fast} (see Fig.~\ref{fig:deltaG}A).
    We would also like to stress that using $z(\textbf{x})$ as CV, we get a smoothly varying free energy surface as compared to the one obtained by projecting the data on $q(\textbf{x})$ (see Supplementary Figure~\ref{sup_fig:chignolin_fes_z_q}). 
    
    In Ref.~\citenum{kang2024computing}, we have shown that the properties of the transition state ensemble could be analyzed in great and illuminating detail with a combination of k-medoid clustering~\cite{Schubert2022kmedoids} and feature relevance analysis.~\cite{pizarroso2022sensitivity}
    Here, as we get a thorough sampling of the entire reaction path, we extend this analysis to understand the full folding mechanism.
    
    To do so, we take advantage of the fact that $q(\textbf{x})$, or equivalently, for simplicity, the smoother $z(\textbf{x})$, provide a natural ordering in the configurations sampled, that can be used to order them as one goes from $q(\textbf{x}\in A)=0$ to $q(\textbf{x}\in B)=1$.
    Thus, we sliced the whole trajectory into 10 bins along the whole $z$ range (that is, $-6.5<z<6.5$) and into 10 more focused on the TS region (that is, $-1.5<z<1.5$), and for each, we identified the most representative configurations through the k-medoid analysis.~\cite{Schubert2022kmedoids}
    Then, we performed a LASSO-based~\cite{novelli2022lasso} feature relevance analysis to identify which descriptors were more relevant for each bin, revealing that the orientation of the side chains does play a role in the formation of the hydrogen bond network that holds the protein together and regulate steric effects.
    In Figure~\ref{fig:chignolin}, we show the superimposition of several configurations belonging to the most representative bins across the (un)folding process.
    
    A first general observation is that, as the folding proceeds, the entropy within the bins decreases, and the corresponding configurations are more similar to each other.
    Interestingly, an explanation can be found guided by the LASSO results (see Supplementary Table~\ref{sup_tab:chignolin_lasso}), which focus the attention on the sidechains of residues Tyr2 and Trp9 in the folded configurations (see Figure~\ref{fig:chignolin} 1 and 2).
    Indeed, moving from frame 1 to 2, the stable folded structure starts to be dismantled due to the misalignment of the side groups of Tyr2 and Trp9, which destroys the hydrophobic $\pi$-$\pi$ interaction between their aromatic rings and owing to the rotation of the side chain of Tyr1, which induces the breaking of the H-bond that locks the end of the main chain.
    
    Once these stabilizing interactions are removed, the whole protein becomes more mobile and can evolve toward the transition state.
    To provide a more detailed description of this crucial region, we report configurations obtained with a denser binning in the range $-1.5<z<1.5$ (3, 4, and 5).
    Using k-medoid clustering, we then identified two clusters within each bin, corresponding to two pathways for the (un)folding of chignolin.
    As a general feature, all the TSE configurations display the bend in the main chain, but the two paths differ by the order in which the stabilizing H-bonds involving residues Asp3, Thr6, and Thr8 are broken (up and down labels in Figure~\ref{fig:chignolin}, respectively).
    These observations are consistent with the result obtained in our previous work~\cite{kang2024computing} and we also observed the TS$^{\text{up}}$ path, which is characterized by the breaking of the upper H-bonds between Asp3 and Thr6, to be dominant over the TS$^{\text{down}}$ path, in which the process starts from the bottom by breaking the H-bonds between Asp3 and Thr8. 
    Again, guided by the results of the LASSO analysis, one can notice that the TS$^{\text{up}}$ path is characterized by the rotation of the side chains of Tyr2 and Trp9 involved in the hydrophobic stacking, whereas the TS$^{\text{down}}$ path mostly involves the main chain arrangements that define the hairpin bend.    
    
    Finally, once at least part of the H-bond network is released, the whole structure of the protein is not stable anymore and thus progressively unfolds, completely losing the characteristic hairpin structure (6 and 7).
    
\begin{figure*}[t!]
    \centering
    \includegraphics[width=0.8\linewidth]{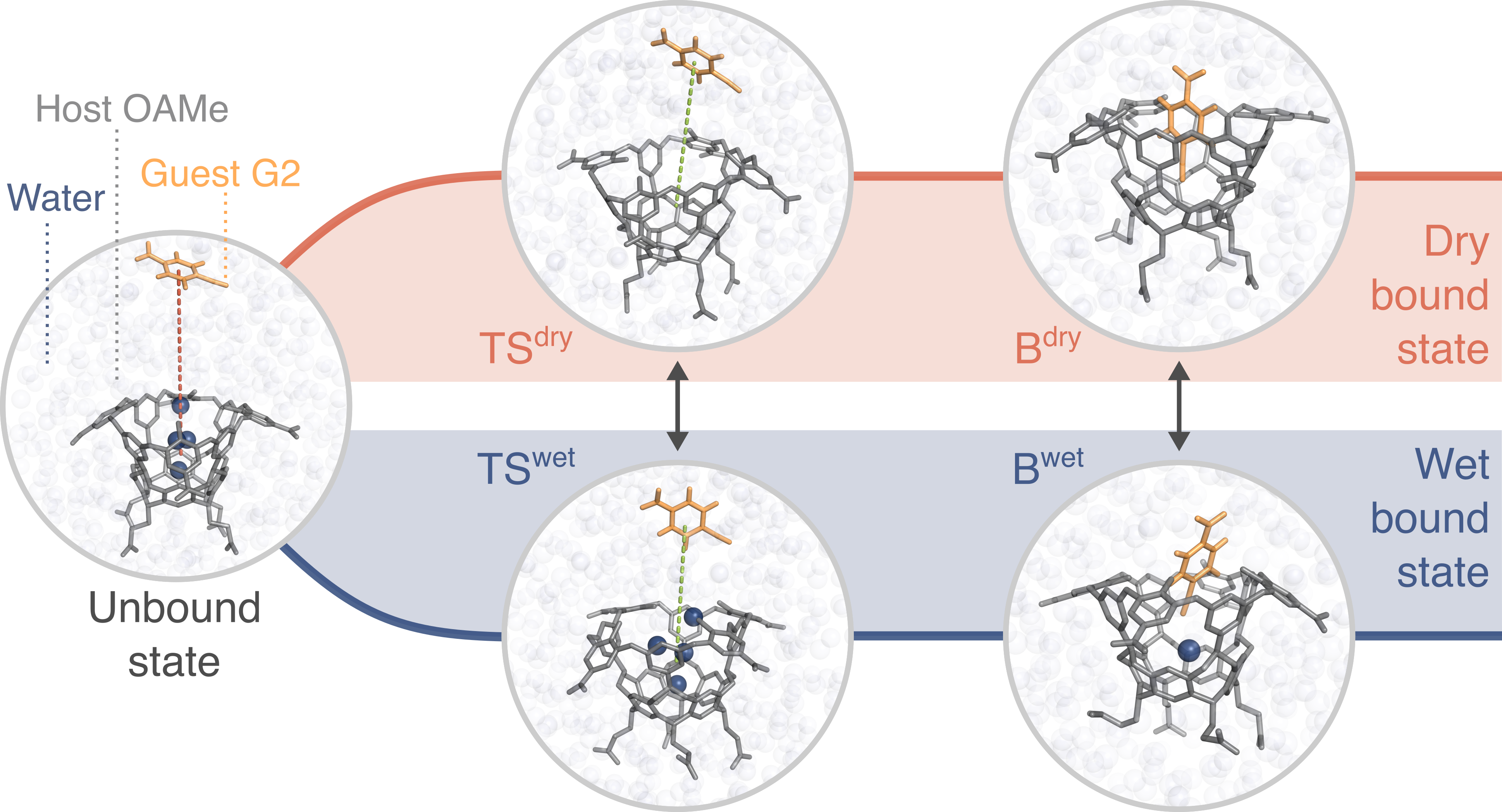}
    \caption{\textbf{OAMe - G2 binding.} Schematic representation of the different possible metastable and transition states found in the binding process of the G2 ligand (orange) to the OAMe host molecule (grey) in water.
    The water molecules are represented as blue spheres, depicted in transparency when outside the binding cavity and in solid color when inside.
    The distance between the host and guest centroids is shown as a dashed line.
    Two main paths from the unbound state to the bound one are classified based on the presence of water inside the pocket as \emph{dry} and \emph{wet}, highlighted in red and blue, respectively.
    Nonetheless, transitions between the transition states TS$^{\text{dry}}$ and TS$^{\text{wet}}$ and the bound states B$^{\text{dry}}$ and B$^{\text{wet}}$ are also observed, as indicated by the black double arrows.
    }
    \label{fig:calixarene}
\end{figure*}

\subsection{Calixarene: Host-Guest systems}

    The final test for our method is the study of the interaction of the G2 ligand with an octa-acid calixarene host (OAMe) from the SAMPL5 challenge,~\cite{yin2017overview} which, although simple, displays some of the key features of the biologically relevant protein-ligand systems. 
    
    Previous work~\cite{rizzi2021role} has studied in detail this system by applying OPES on two independent CVs and using a funnel restraint to limit the unbound state diffusion.~\cite{limongelli2013funnel}
    One CV was the projection of the ligand center of mass on the binding axis ($h$), and the other was a machine-learning CV based on a classification criterion~\cite{bonati2020data} that accounted for the role of water in the process, as it took as inputs the water coordination numbers of a curated set of points along $h$ and on the ligand molecule.
    Here, as inputs for a single committor model, we use both types of information to obtain a simple and complete description of the whole process.
    More in detail, we used the same water coordination numbers and, rather than the $h$ projection, a set of distances between the ligand molecule and the binding pocket to have more flexibility in also accounting for the relative orientation of the ligand.
    
\begin{figure}[h!]
    \centering
    \includegraphics[width=1\linewidth]{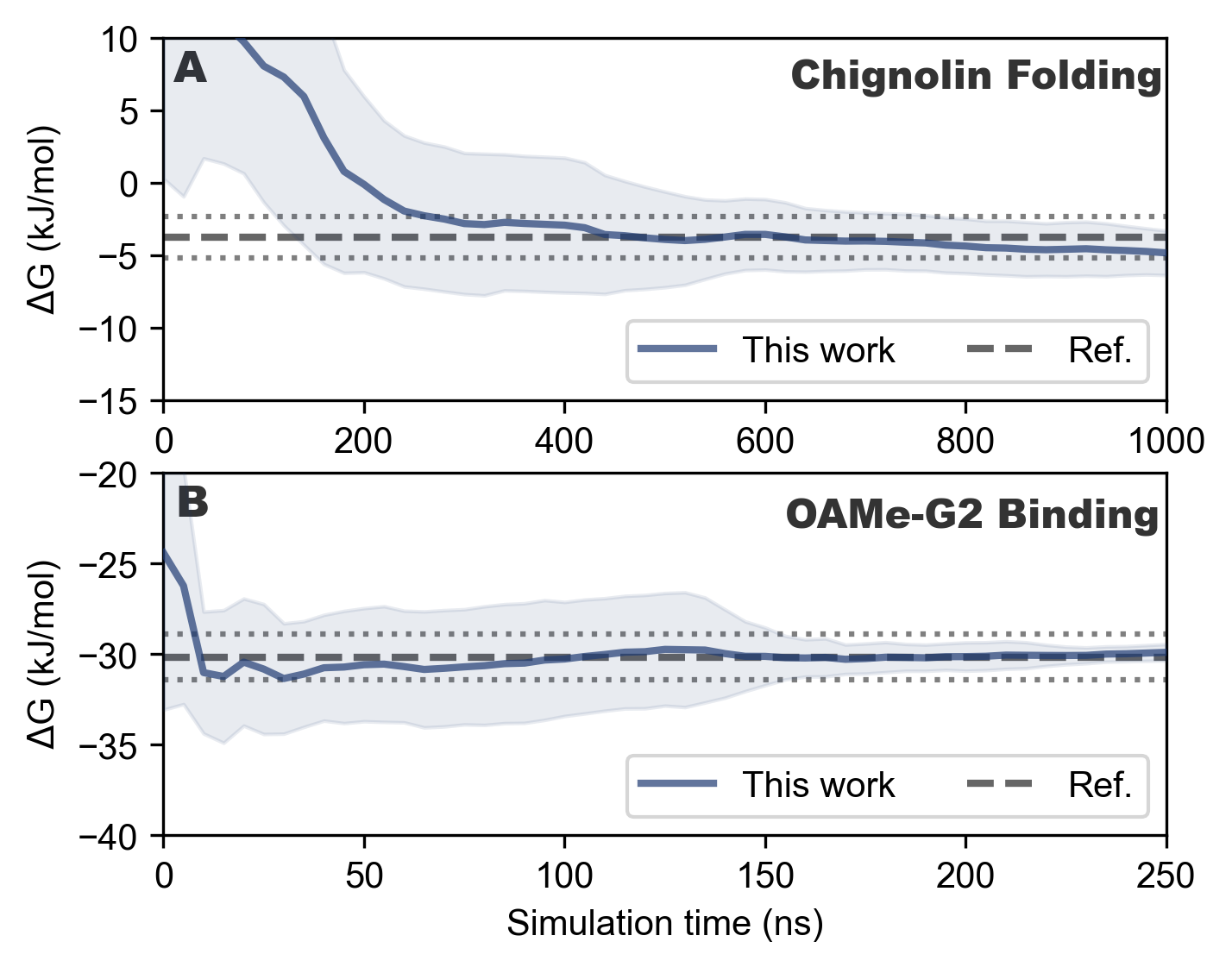}
    \caption{\textbf{Convergence of $\boldsymbol{\Delta} \textbf{G}$} Convergence with simulation time of the estimate for the folding energy of chignolin (\textbf{A}) and the binding energy of the G2 ligand to the OAMe octa-acid guest (\textbf{B}). 
    The average estimates from independent simulations for each system \change{(3 for chignolin, 4 for OAMe-G2)} are reported as a blue solid line, whereas the uncertainty, computed as the standard deviation over the three replicas, is depicted as a shaded blue region.
    The reference values are provided as grey dashed lines, and the $0.5\,k_BT$ interval around the reference is marked by grey dotted lines.
    For chignolin, that is the unbiased estimate from Ref.~\citenum{lindorff2011fast}, for the calixarene, an estimate obtained using the enhanced sampling setup of Ref.~\citenum{rizzi2021role}.}
    \label{fig:deltaG}
\end{figure}

    As a starting point for our procedure, we ran unbiased simulations in the bound state (B$^{\text{dry}}$) and the unbound state (U) only (see Figure~\ref{fig:calixarene}), and within a few iterations, we converged the committor.
    At that point, we could obtain the FES for the (un)binding process (see Supplementary Figure~\ref{sup_fig:Calixarene_2DFES}) and, after proper correction for the funnel restraint (see Supplementary Section~\ref{sup_sec:funnel_restraint}), we could also get an estimate of the binding free energy
    in good agreement with results obtained with the same setup and the collective variables presented in the literature~\cite{rizzi2021role} (see Fig.~\ref{fig:deltaG}B).

    However, the 2D FES clearly shows the presence of an intermediate state close to the bound state, which suggests that a mere two-state scenario does not apply to this process, as it often happens when dealing with complex systems.
    Such an intermediate state is the same semi-bound state observed in Ref.\citenum{rizzi2021role} in which a water molecule in the binding pocket prevents the ligand from correctly fitting.
    Not surprisingly, such a \emph{wet bound state} (B$^{\text{wet}}$) is associated with a predicted committor value of $q\approx1$, which is more than reasonable as it is much more similar to the true \emph{dry bound state} (B$^{\text{dry}}$) than to the unbound one.
    More specifically, one can find that the B$^{\text{wet}}$ state is more likely to transform into state B$^{\text{dry}}$, as the FES clearly indicates that the largest barrier is related to the transition to U that represents the true rate-limiting step for the process.  

    
    Moreover, our committor-based simulations allowed us to characterize different reaction channels in a single calculation, which we depict in Figure~\ref{fig:calixarene} and can be identified by visualizing the projection of the Kolmogorov ensemble in the plane defined by $h$ and the water coordination number inside the cavity $V_2$ (see Supplementary Figure~\ref{sup_fig:Calixarene_2D_K}).
    As one could expect, the TS configurations are mostly determined by the distance of the ligand from the host molecule, as proved also by the feature relevance analysis (see Supplementary Figure~\ref{sup_fig:calixarene_feature_relevance}), which is sensibly shortened with respect to the fully unbound state.
    However, from $p_K$, it can also be seen that there are two contributions to the TSE, differing in the behavior of water and, similar to what is discussed in Section~\ref{sec:double_path}, one dominates over the other.

    The dominant paths pass through a TS with water inside the cavity (TS$^{\text{wet}}$), evolving then to the B$^{\text{wet}}$ state and eventually to the B$^{\text{dry}}$.
    In the other case, the TS already does not present water inside the pocket (TS$^{\text{dry}}$), thus allowing for a direct evolution to the proper B$^{\text{dry}}$ state.

\section{Discussion}
    In this work, we have presented a new enhanced sampling strategy for addressing the rare event problem that combines accurate free energy calculations with extensive sampling of transition states.
    Building on our recent work,~\cite{kang2024computing} our approach is based on a semi-automatic self-consistent iterative procedure for the determination of the committor function using the variational principle it obeys.
    A crucial addition to our previous workflow is the definition of a committor-related collective variable that is used not only to drive metadynamics-like calculations but also as an analysis tool for the interpretation of the results to obtain precious physical insights.
    These improvements lead to much more efficient and accurate learning of the committor function and also to accurate estimates of free energy profiles.
    
    Nonetheless, there is still room for improvement.
    For example, it would be highly desirable to remove the need for physical descriptors, thus providing an even more general approach, for instance, by exploiting more sophisticated models based on graph neural networks that recently have been making their way also in the enhanced sampling field.~\cite{zhang2024descriptorsfree, dietrich2024graph, zou2024graph}
    On the other hand, from the theoretical point of view, improvements could be obtained by eventually extending the ideas behind the committor function to a more general framework that could help tackle more efficiently complex systems characterized by rugged energetic landscapes and multiple intermediates.
    \change{This should also include the possibility of estimating kinetic rates directly from the committor as prescribed in transition path theory}.\cite{weinan2010transition}
    \change{In addition, it would also be interesting to investigate the applicability and possible advantages of employing our committor-related CV with other enhanced sampling methods in which directly using the committor was problematic.~\cite{rotskoff2022active, aristoff2023weighted}}

    We believe that this approach opens the way to a new era of enhanced sampling, in which reactive processes are studied through the lens of probability distributions, which the committor encodes.
    Indeed, more and more studies on complex systems are recognizing the importance of embracing the statistical complexity of reactive events in fields as different as catalysis~\cite{wang2015dynamic, wang2016oxidation, yang2023reactant, tian2023dynamically, perego2024dynamics, tripathi2024poisoning} and biophysics.~\cite{lindorff2011fast, dill2012folding, sztain2021glycan}
    This further convinces us that a probabilistic description is the natural way to accurately study and analyze in-depth complex reactive processes.

\section{Methods}  
\label{sec:methods}
    In the following sections, we first provide a general background on the committor and the variational principle it obeys, and we summarize the concepts of the Kolmogorov bias and the related ensemble we have introduced in Ref.\citenum{kang2024computing}  (Section~\ref{sec:background}).
    We then discuss the improved iterative protocol for the self-consistent determination of the committor function  (Section~\ref{sec:learning_committor}) and its use to converge free energy calculations  (Section~\ref{sec:committor_cv}), highlighting the advancements from our previous work.~\cite{kang2024computing}
    In the last section, we schematically summarize the whole iterative procedure to help contextualize all the details  (Section~\ref{sec:summary}).
    
\subsection{Background}
    \label{sec:background}
    \paragraphtitle{The committor function}
        The committor function $q(\textbf{x})$ was introduced by Kolmogorov for the study of rare events involving transitions between two metastable states $A$ and $B$.~\cite{kolmogoroff1931analytischen}
        Such a function is defined as the probability that a trajectory started at configuration $\textbf{x}$ ends up in $B$ before first passing by $A$~\cite{weinan2010transition} and is thus believed to be the best possible formal description for a rare event process.~\cite{berezhkovskii2005one,li2014recent,ma2005automatic,he2022committor}
        Unfortunately, determining the committor function is a far from trivial challenge.
    
        In the transition path sampling literature,~\cite{Dellago2006, jung2023machine, bolhuis2002throwing} the committor value for a specific configuration $\textbf{\tilde{x}}$ is evaluated performing the so-called committor analysis, in which many independent trajectories are started from $\textbf{\tilde{x}}$ to statistically evaluate $q(\textbf{\tilde{x}})$ based on the outcome of such simulations. 
        Despite its appealing simplicity, this approach presents some serious drawbacks that limit its viability. 
        For example, its computational cost can easily become very heavy as the number of selected $\textbf{\tilde{x}}$ configurations grows, also because many simulations have to be performed for each $\textbf{\tilde{x}}$ to obtain meaningful statistics.
        More critically, it is also known that when one tackles more complex systems, the results of such an analysis can be strongly affected by the criteria used to determine whether a trajectory is committed to either basin A or B.~\cite{lazzeri2023committor}

        However, Kolmogorov also showed a theoretical approach to compute the committor function for a generic system governed by an interatomic potential $U(\textbf{x})$ at inverse temperature $\beta$. This is rigorously derived under the hypothesis of overdamped Langevin dynamics, but with a few reasonable assumptions, it can also be extended to the general case.~\cite{weinan2010transition}
        Conveniently, such an approach can be equivalently formulated as a differential or a variational problem.

        The first one is based on a set of partial differential equations that obey the boundary conditions $q(\textbf{x}_A) = 0$ and $q(\textbf{x}_B) = 1$ on the starting and final points of the transition $\textbf{x}_A$ and $\textbf{x}_B$. 
        Unfortunately, a practical application of this approach is unfeasible, as its equations are impossible to solve if not for a few trivial low-dimensional toy models.
        The second approach, on the other hand, finds an equivalent solution following a variational principle that obeys the same boundary conditions.
        In practice, this amounts to minimizing the functional of the committor
            \begin{equation}
                K[q(\textbf{x})]=\frac{1}{Z} \int d\textbf{x} |\nabla_\textbf{u} q(\textbf{x})|^2 e^{-\beta U(\textbf{x})}
                \label{eq:variational_functional_extended}
            \end{equation}
        where $Z=\int d\textbf{x} e^{-\beta U(\textbf{x})}$ is the Boltzmann partition function and $\nabla_\textbf{u}$ denotes the gradient with respect to the mass-weighted coordinates $\textbf{u}_i^j = \sqrt{m^j} \textbf{x}_i^j$, in which $m^j$ is the mass of atoms of type $j$. 
        
        As it often happens with variational problems, applying this second approach is less demanding, making it a viable option.
        The functional in Eq.~\ref{eq:variational_functional_extended} can indeed be written equivalently as the expectation value of the square modulus of the committor gradients $|\nabla_\textbf{u} q(\textbf x)|^2$, which is related to the reaction rate $\nu_R$, and the average is over the Boltzmann ensemble
            \begin{equation}
                K[q(\textbf{x})] = \langle |\nabla_\textbf{u} q(\textbf x)|^2 \rangle _{U(\textbf{x})}  
                \label{eq:variational_functional_sampling}
            \end{equation}
        and thus, it can be estimated through sampling.
        
        However, even this approach still presents some serious challenges.
        In a rare event scenario, the committor is indeed almost constant in the metastable basins, that is, $q\simeq0$ in A and $q\simeq1$ in B, and smoothly interpolates between these two values in the transition region.
        As a consequence, the gradients that appear in Eq.~\ref{eq:variational_functional_sampling} are strongly localized on the transition region, which, unfortunately, is exactly the region that is difficult to sample in simulations.
        To bypass this sampling difficulty, in Ref.~\citenum{kang2024computing}, we introduced the Kolmogorov bias potential, a new enhanced sampling protocol that, by design, focuses the sampling on the TS, which we introduce in the following section.
    
    \paragraphtitle{Sampling the Kolmogorov ensemble}
        In a rare event scenario, metastable states are separated by large free energy barriers that hinder the transitions between them.
        The set of high-energy configurations that are found at the top of the saddle points associated with such barriers is usually referred to as the transition state ensemble (TSE).
        Knowledge of the TSE is highly desired as it can provide precious insights into reaction mechanisms and rates, but unfortunately, sampling the TSE with MD simulations is typically an issue.
        
        Such configurations are indeed completely out of reach in standard MD and are difficult to sample even when conventional metadynamics-like enhanced sampling methods are applied.~\cite{laio2002escaping, invernizzi2020rethinking}
        Indeed, in that case, the bias potential added to the system aims at filling the metastable basins to promote transitions between them by lowering the relative free energy barriers.
        However, in the biased energy landscape, the TS region still remains a local maximum whose sampling is thus unfavored compared to the metastable basins (see Figure~\ref{fig:NN_z_q}).
        In addition, the quality of the TSE sampling with such calculations also depends on the quality of the used CV. 
        
        Ideally, if one wants to study the TS, one would like to overturn this scenario, thus building a bias potential that, by design, focuses the sampling on the TS rather than wasting precious simulation time in the metastable basins.
        However, this sounds like a chicken-and-egg problem, as to build such a bias, we need to locate the TS, which is the very reason why we want to build the bias.
        Remarkably, a solution to this dilemma can be obtained by taking inspiration from the committor itself.
        
        In Ref.~\cite{kang2024computing}, we introduced the Kolmogorov TS-oriented bias potential
            \begin{equation}
                V_K(\textbf{x})=-\frac{\lambda}{\beta}\log (|\nabla q(\textbf{x})|^2)
                \label{eq:kolmogorov_bias}
            \end{equation}
        where $\lambda \simeq 1$ is a parameter used to scale the magnitude of the bias. 
        $V_K$ depends on the gradients of the committor function, which, as discussed in the previous section, are localized on the TS by construction.
        As a consequence, in the resulting biased energy landscape $U_K(\textbf{x}) = U(\textbf{x}) + V_K(\textbf{x})$, the TS region is stabilized over the rest of the phase space, favoring its sampling (see Figure~\ref{fig:NN_z_q}).

        Starting from $U_K$, we can define the Kolmogorov probability distribution
            \begin{equation}
                p_K(\textbf x) = 
                \frac{e^{-\beta U_K(\textbf{x}) }}{Z_K} \quad\text{with}\quad Z_K=\int e^{-\beta U_K(\textbf{x}) }d\textbf{x}
                \label{eq:kolmogorov_distribution}
            \end{equation}
        in which only those configurations that actually contribute to the reaction rate have a non-negligible contribution.
        Indeed, in such a distribution, the relative weight of the configurations is determined not only by the probability that state $\textbf{x}$ is sampled, which is proportional to the Boltzmann weight $e^{-\beta U(\textbf{x})}$, but also by the contribution that successful reactive trajectories passing by $\textbf{x}$ bring to the reaction rate, which is expressed by the $e^{-\beta V_K(\textbf{x})} = e^{-\log(|\nabla q(\textbf{x})|^2)} = |\nabla q(\textbf{x})|^2$ term. 
        For this reason, we proposed using the $p_K(\textbf{x})$ to formalize an improved definition of the TSE, thus extending the conventional definition that includes all those configurations found on the $q\simeq0.5$ isosurface, regardless of their actual probability of being sampled in the true Boltzmann distribution.

    \paragraphtitle{Self-consistent approach for the committor}
        In Ref.~\citenum{kang2024computing}, we have shown how the variational principle and the sampling of the Kolmogorov ensemble introduced above can be combined in a self-consistent iterative procedure to obtain the committor function with the help of machine learning.
        In the iterative procedure, we alternate training and sampling cycles until convergence is reached. 
        The first are aimed at improving a neural-network-based variational parametrization of the committor, and the second generate new and progressively better data to learn from under the effect of the TS-oriented Kolmogorov bias.
        In the following sections, we present an improved version of the procedure that not only allows for a faster convergence through the iterations but also the calculation of the free energy profile of the process.

    \subsection{Machine learning the committor function}
    \label{sec:learning_committor}
        In the method presented in Ref.~\citenum{kang2024computing}, the committor function is represented as a neural network (NN) $q_\theta(\textbf{x}) = q_\theta(\textbf{d}(\textbf{x}))$ with learnable parameters $\theta$ that takes as inputs a set of physical descriptors $\textbf{d}(\textbf{x})$ (see Figure~\ref{fig:NN_z_q}).
        Such descriptors are functions of the atomic coordinates, for instance, distances or coordination numbers\change{, and as for the case of machine-learning CVs based on the same framework,~\cite{bonati2023mlcolvar, trizio2024advanced} they should be chosen so that they can describe the relevant physics of the systems when considered collectively. In addition, in Ref.\citenum{kang2024computing}, we have shown how the quality of different descriptor sets can be evaluated using the lowest achievable variational bound as a figure of merit.}
        
        The optimization of $q_\theta^n (\textbf{x})$, at iteration $n$, is based on the minimization of an objective function $L$ that formalizes the variational principle of Eq.~\ref{eq:variational_functional_sampling} and its boundary conditions.
        In fact, the loss function presents two terms $L = L_v + \alpha L_b$, relatively scaled by the hyperparameter $\alpha$.
        The $L_v$ term derives from the variational principle and reads
            \begin{equation}
                L_v = \frac{1}{N^n} \sum_i^{N^n}  w_i |\nabla_\textbf{u} q(\textbf{x}_i)|^2
                \label{eq:loss_variational}
            \end{equation}
        where the weights $w_i$ reweigh each configuration $\textbf{x}_i$ to it natural Boltzmann weight. 
        The $L_b$ term enforces the boundary conditions $\textbf{x}_A=0$ and $\textbf{x}_B=1$ as
            \begin{equation}
                L_b = \frac{1}{N_A}\sum_{i \in A}^{N_A} (q(\textbf{x}_i))^2 + \frac{1}{N_B}\sum_{i \in B}^{N_B} (q(\textbf{x}_i) - 1)^2
                \label{eq:loss_boundary}
            \end{equation}
        In Ref.~\citenum{kang2024computing}, the $L_b$ term was evaluated on a labeled dataset of $N_A$ and $N_B$ configurations collected at the first iteration by running short, unbiased MD simulations in the two basins.
        On the other hand, the $L_v$ term was computed on the whole dataset available, thus including both the initial unbiased and biased $N^n$ configurations from later iterations. 
        This was helpful, as the sampling cycles under the action of the $V_K$ bias were focused on the TS region but not optimal, as the unbiased configurations were incorrectly weighted in the dataset.
        This issue made it necessary to perform more iterations or correct the corresponding weights (that is, $w_i$ when $i\in A,B$) by an estimate of the free energy difference between the two basins, which is not always readily available.

        Our new approach gets rid of this slowdown, as at each sampling iteration, we can effectively visit not only the TS region but also both metastable states (see Section~\ref{sec:committor_cv}).
        In this way, the new configurations cover the whole phase space, and after proper reweighting, the weights $w_i$ automatically follow the real Boltzmann distribution.
        As a consequence, we can thus avoid evaluating the $L_v$ term on the unbiased labeled configurations, making the whole procedure more precise and faster. 
        It should also be noted that, in the first iteration, when only unbiased data are available, the contribution from the $L_v$ term was almost null also in the original formulation due to the localization of the gradients $|\nabla q(\textbf{x}) |^2$ discussed in Section~\ref{sec:background}.
        A second consequence of the higher quality of the data thus generated is a more data-efficient protocol.
        Indeed, as soon as the enhanced sampling becomes effective (that is, multiple transitions can be observed), to optimize the parameters $\theta$ at each iteration, it is sufficient to use configurations and weights generated in the previous iteration or, at most, from the previous two.
        
\subsection{Extensive sampling along a committor-based CV}
    \label{sec:committor_cv}
        As already anticipated in the introduction, we bring two conceptual improvements to the original recipe of Ref.~\citenum{kang2024computing}.
        First, we define a physics-informed collective variable (CV) $z(\textbf{x})$ that is based on the committor and is effective when used with metadynamics-like enhanced sampling approaches.
        Second, we improve such approaches by combining them with the Kolmogorov bias potential to favor the TS sampling.

        Even if the committor has often been thought to be an ideal CV,~\cite{berezhkovskii2005one,li2014recent,ma2005automatic,he2022committor} we already pointed out in the introduction that its straightforward application in an enhanced sampling context is far from optimal due to the high degeneracy of the metastable states along such a variable, as $q(\textbf{x_A})\simeq0$ and $q(\textbf{x_B})\simeq1$, and to its sharp behavior in the transition region.
        This would suggest that a smoother modification of the committor might be an effective CV.
        A practical way of simply and economically achieving this result comes at zero cost from the NN-based parametrization of the committor proposed in Ref.~\citenum{kang2024computing}.
        There, as we have seen in Section~\ref{sec:learning_committor} and depicted in Figure~\ref{fig:NN_z_q}, the committor is represented as a NN, and a sharp sigmoid activation function $\sigma$ is applied to the output layer of such a network to enforce the correct functional form to the learned function, that is, $q(\textbf{x}) = \sigma(z(\textbf{x}))$.
        It is thus wiser to employ as a CV, not the activated final committor function, but rather the output $z(\textbf{x})$ of the NN.
        In fact, this still preserves the physical information encoded in the committor function, but it is also suitable to be used in an enhanced sampling context as it still has finite gradients in the metastable regions, and it is not as sharp in the TS region.

        As one can expect, similar numerical issues could also affect the bias potential $V_{K}$ that, in Ref.~\citenum{kang2024computing}, was computed as a function of the gradients of the committor function $\nabla q(\textbf{x})$.
        In fact, this choice is not optimal in practice as, due to the sharp behavior of $q(\textbf{x})$, such a computation can easily lead to extremely small values below machine precision in the basins where $q(\textbf{x})$ is flat (see Supplementary Figure~\ref{sup_fig:numerical_issues}), thus making the bias almost ineffective there.
        Conveniently, by taking advantage of the change of variable $q \rightarrow z$ and with the help of some elementary algebra (see Supplementary Section~\ref{sup_sec:bias_z}), one can bypass this issue, writing the bias potential as a function of the smoother and more numerically stable gradients of $z$.
       
        Like any CV, $z$ can be used, in principle, with any CV-based method. 
        However, here, we opt for the On-the-fly Probability Enhanced Sampling (OPES) method~\cite{invernizzi2020rethinking, trizio2024advanced}, a recent evolution of metadynamics~\cite{laio2002escaping} that arguably represents the state of the art in this field.
        We already pointed out in Ref.~\citenum{kang2024computing} that, in metadynamics-like frameworks, the metastable basins are \emph{filled} with an external bias potential to reduce the height of free energy barriers and promote transitions, but the TS-region is still found at a local maximum, whose sampling is thus unfavored (see Figure~\ref{fig:NN_z_q}).
        
        Here, we propose to remove this limitation by adding a Kolmogorov bias $V_K$ to such a biasing scheme. 
        In fact, this will stabilize the TS, turning it into a minimum that will eventually be filled by the original OPES bias $V_{\text{OPES}}$ alongside the true metastable states. 
        In practice, this amounts to driving the system with an effective bias potential $V_{\text{eff}} = V_K + V_{\text{OPES}}$.
        In this way, one can promote frequent reactive events while increasing the time spent in the usually elusive transition region. 
        As a result, one can thus fully characterize the rare event by harvesting in the same simulation a wealth of precious TSE configurations and obtaining an accurate estimate of the free energy surface (FES).
        This can be recovered with a simple reweighting by the total effective bias potential according to 
        \begin{equation}
            w^n_i = \frac{e^{\beta V_{\text{eff}}^n(\textbf{x}_i)}}{\langle e^{\beta V_{\text{eff}}^n(\textbf{x})} \rangle_{U^n_{\text{eff}}} }
        \end{equation}
        which benefits from the fast-converging OPES framework and the static nature of the Kolmogorov bias. 
        
        As one can expect, the improved sampling of this new approach is also beneficial to the iterative procedure as it greatly simplifies the dataset construction, exploiting the best from the two biasing schemes.
        \change{From the Kolmogorov bias, it inherits an extensive sampling of the TSE, which makes it possible to apply accurately the variational principle (see Section~\ref{sec:background}).}
        \change{On the other hand,} from OPES, \change{we obtain} a fast-converging bias potential that allows sampling of the relevant part of the phase space.~\cite{invernizzi2022exploration, trizio2024advanced}
        This facilitates the reweighting, thus guaranteeing a more accurate estimate of the Boltzmann weights and substantially reduces the number of iterations needed for convergence (see Section~\ref{sec:learning_committor}).
        \change{Moreover, integrating eventual additional CVs in the OPES biasing scheme is simple, which could help to further improve the simulation efficiency in some complex cases, for instance, biasing orthogonal degrees of freedom that distinguish different paths if present.}

    \subsection{Summary of self-consistent iterative procedure}
    \label{sec:summary}

        Here, we summarize schematically the whole iterative procedure, in which we alternate training and sampling cycles.
            \begin{itemize}
                \item \textbf{Step 1:} The NN-based committor   $q_\theta^n(\textbf{x})$ at iteration $n$ is optimized on the dataset of configurations $\mathbf{x}^{n}$ and weights $w_i^{n}$ updated from the previous iterations. 
                To start the procedure, at the first iteration $n=0$, we build such a dataset by running some (short) simulations in the metastable basins and labeling them accordingly.
                The optimization is based on the variational principle of Eq.~\ref{eq:variational_functional_sampling}, minimizing the loss function of Eqs.~\ref{eq:loss_variational} and \ref{eq:loss_boundary} (see Section~\ref{sec:learning_committor}).
                
                \item \textbf{Step 2:} We perform biased simulations under the coupled action of the Kolmogorov and OPES bias potentials, $V^n_K$ and $V^n_{\text{OPES}}$.
                The latter is applied along the committor-related CV $z^n(\textbf{x})$ and drives the sampling across the whole relevant configurational space between the two states, whereas the former guarantees a thorough sampling of the TS region crucial for applying Eq.~\ref{eq:variational_functional_sampling} (see Section~\ref{sec:committor_cv}).
                
            \item \textbf{Step 3:} The dataset is updated with the new sampled configurations, which are reweighted by the effective bias $V_{\text{eff}} = V_K + V_{\text{OPES}}$ according to $w^n_i = \frac{e^{\beta V_{\text{eff}}^n(\textbf{x}_i)}}{\langle e^{\beta V_{\text{eff}}^n(\textbf{x})} \rangle_{U^n_{\text{eff}}} }$. 
                A progressively more accurate estimate of the system's FES is obtained from such reweighted data. 
            \end{itemize}

\section*{Data Availability} \label{sec:code_avail}
    Training and simulation data and inputs are available on \hyperlink{https://github.com/EnricoTrizio/committor\_2.0}{GitHub}~\cite{trizio2025github} and on \hyperlink{https://zenodo.org/records/15089373}{Zenodo}.~\cite{trizio2025dataset} 

\section*{Code Availability} \label{sec:code_avail}
    The code for the training of the NN-based committor model alongside didactic tutorials is available through the open-source \texttt{mlcolvar} library~\cite{bonati2023mlcolvar}, which is the preferred way to access the most updated code. To obtain the results reported in the manuscript, version 1.2.2 was used and a frozen version is also available on \hyperlink{https://zenodo.org/records/15089373}{Zenodo}.~\cite{trizio2025dataset}
    The PLUMED~\cite{tribello2014plumed, plumed2019promoting} interface for the application of the bias is available on \hyperlink{https://github.com/EnricoTrizio/committor\_2.0}{GitHub}~\cite{trizio2025github} and on \hyperlink{https://zenodo.org/records/15089373}{Zenodo}.~\cite{trizio2025dataset}

\begin{acknowledgments}
    The Authors are grateful to Luigi Bonati, Andrea Rizzi, Jintu Zhang, Umberto Raucci, Alice Triveri, and Francesco Mambretti for the useful discussions and feedback about this manuscript. 
\end{acknowledgments}

\section*{Authors contributions}
    All the authors equally contributed to the manuscript by conceptualizing the project, developing the theoretical methodologies, and participating in the writing of the manuscript.
    Specifically, E.T. and P.K. developed the code for the training of the committor function and conducted the computational simulations.

\section*{Competing interests statement}
    The authors declare no competing interests.

\section*{References}
\bibliography{references}

\setcounter{section}{0}
\renewcommand{\thesection}{S\arabic{section}}
\setcounter{equation}{0}
\renewcommand{\theequation}{S\arabic{equation}}
\setcounter{figure}{0}
\renewcommand{\thefigure}{S\arabic{figure}}
\setcounter{table}{0}
\renewcommand{\thetable}{S\arabic{table}}
    
\clearpage
\onecolumngrid

{\Large\normalfont\sffamily\bfseries{{Supporting Information}}}

\setlength{\tabcolsep}{18pt}
\renewcommand{\arraystretch}{1.2}

\setlength{\abovecaptionskip}{0.5pt} 

\section{Numerical stability of Kolmogorov bias potential} \label{sup_sec:bias_z}
\begin{figure}[h!]
    \centering
    \begin{minipage}{0.48\linewidth}
        \includegraphics[width=0.8\linewidth]{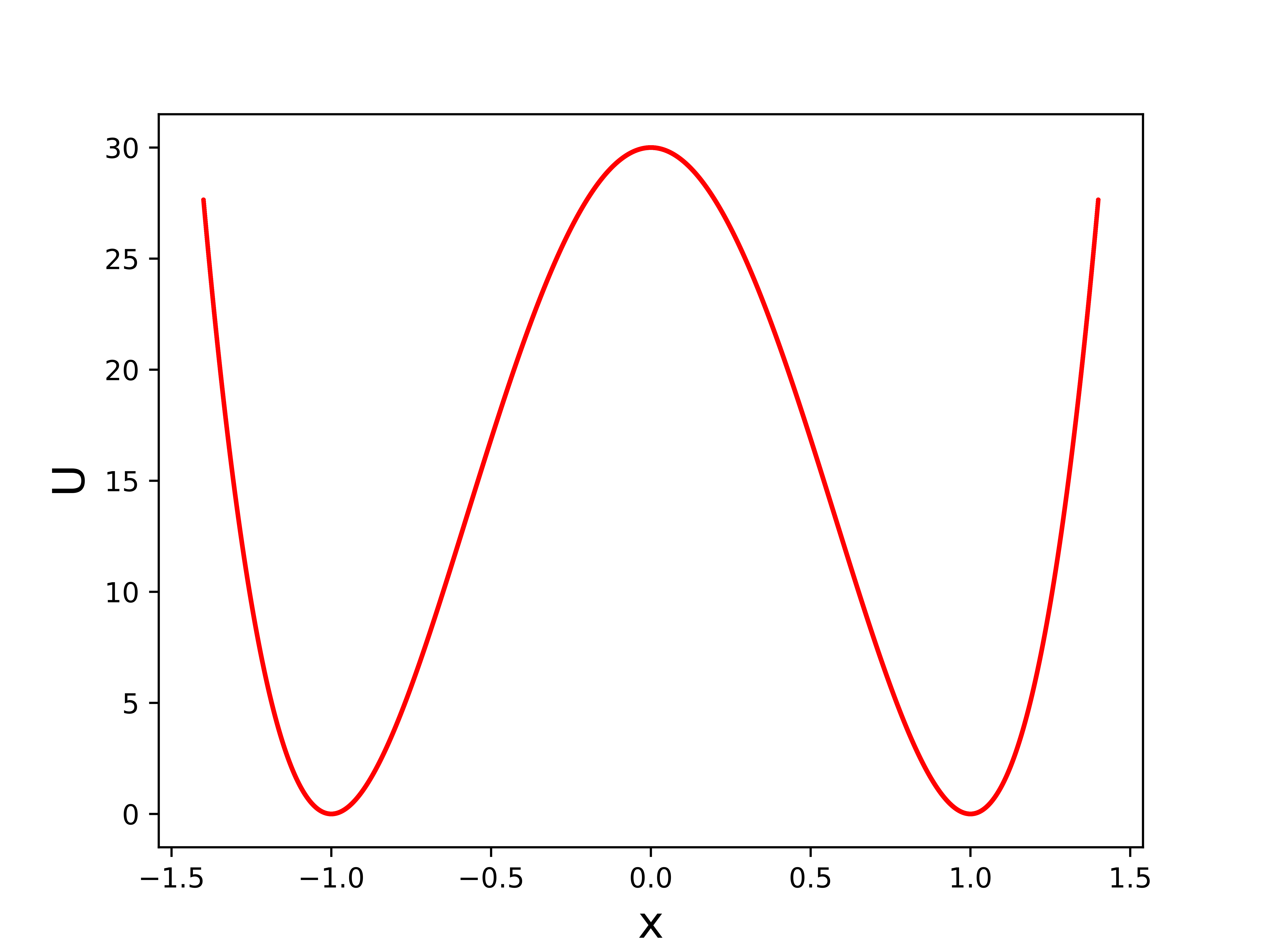}
    \end{minipage}
    \begin{minipage}{0.48\linewidth}
        \includegraphics[width=0.8\linewidth]{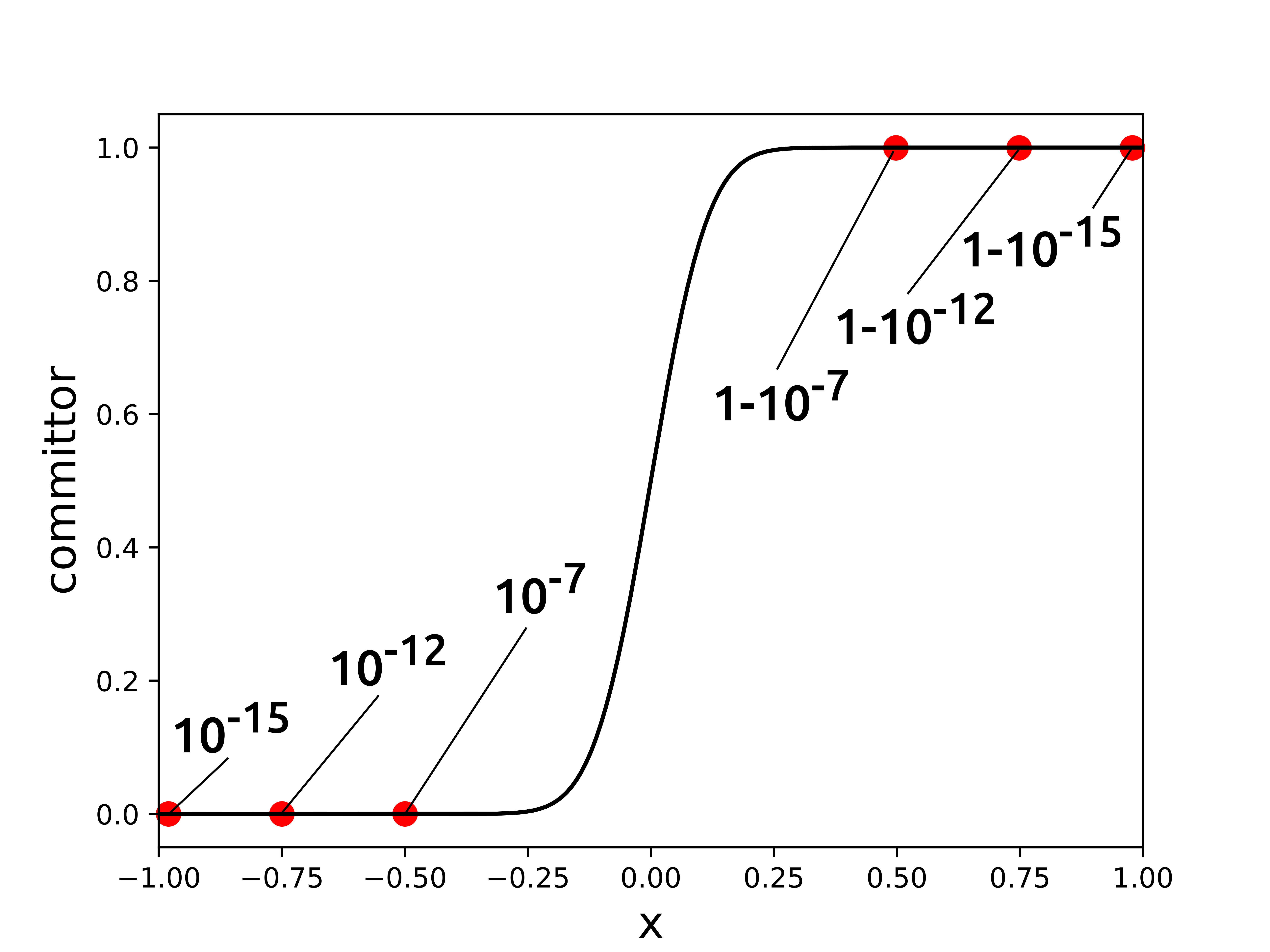}
    \end{minipage}
    \caption{\textbf{Numerical issues of the committor function} Visualization of the numerical issue related to the committor function on a toy double-well potential along one dimension x for which an analytical solution can be obtained.}
    \label{sup_fig:numerical_issues}
\end{figure}

        To improve the numerical stability and efficacy of the Kolmogorov bias, it is convenient to reformulate its calculation (Eq.~\ref{eq:kolmogorov_bias}) in terms of the committor-related $z$ collective variable as it allows lifting the numerical problems that may arise from the direct use of $q$ (see Figure~\ref{sup_fig:numerical_issues}).
        Here, we show that this is formally equivalent to expressing as a functional of $q$.
        In the main text, we shown that $q$ is obtained applying a sigmoid-like switching function $\sigma$ to $z$
            \begin{equation}
                q(\textbf{x}) = \sigma(z(\textbf{x}))
                \quad\text{with}\quad
                \sigma(z) = \frac{1}{1+e^{-pz}}
            \end{equation}
        where we set $p=3$.
        Thus, using the chain rule, we can write
            \begin{equation}  
                \frac{\partial{q(\textbf{x})}}{\partial x} = 
                \frac{\partial{\sigma(z)}}{\partial z}\frac{\partial{z}}{\partial x}
            \end{equation} 
        After some elementary algebra, we can equivalently rewrite the bias as
            \begin{equation}  
                V_{K} = \frac{1}{\beta} \big[ \log(|\nabla z|^2) -4\log({1+e^{-pz}} ) -2pz \big] + c
            \end{equation}  
        where $c$ is a constant that thus will not affect free energy calculations.

        Such a bias is equivalent to the original one but has the advantage that $\frac {\partial z}{\partial x}$ has a much higher numerical resolution in the metastable basins, where $q\sim0$ and $q\sim1$.

\clearpage
\section{M\"uller-Brown potential - Additional information}
    \subsection{Computational details}
        \paragraphtitle{Simulations details}
            The M\"uller-Brown potential energy surface, $U(x, y)$, is defined as a function of the Cartesian coordinates $x$ and $y$ 
                \begin{equation}
                    U(x,y) = -k\sum_{i=1}^4 d_i e^{a_i(x-x_i) + b_i (x-x_i)(y-y_i) + c_i(y-y_i)}
                    \label{seq:muller_formula}
                \end{equation}
            where the constants take the following values, k = 0.15, [$d_1$, $d_2$,  $d_3$,  $d_4$ ]= [-200, -100, -170, 15], [$a_1$, $a_2$,  $a_3$,  $a_4$] = [-1, -1, -6.5, 0.7], [$b_1$, $b_2$,  $b_3$,  $b_4$] = [0, 0, 11, 0.6], [$c_1$, $c_2$,  $c_3$,  $c_4$] = [-10, -10, -6.5, 0.7], [$x_1$, $x_2$,  $x_3$,  $x_4$] = [1, 0, -0.5, -1] and  [$y_1$, $y_2$,  $y_3$,  $y_4$]  = [0, 0.5, 1.5, 1].
            
            The simulations of the diffusion of an ideal particle of mass 1 have been performed using Langevin dynamics based on the Bussi-Parrinello algorithm~\cite{bussi2007accurate} as implemented in the \texttt{ves\_md\_linearexpansion}~\cite{valsson2014variational} module of PLUMED.~\cite{tribello2014plumed, plumed2019promoting}
            The damping constant in the Langevin equation was set to  10/time-unit. 
            The time unit was defined arbitrarily and corresponds to 200 timesteps and natural units ($k_BT = 1$) were used in all the calculations.

        \paragraphtitle{Committor model training details}
            To model the committor function $q_\theta(\textbf{x})$ at each iteration, we used the x and y Cartesian coordinates of the diffusing particle as inputs of a neural network (NN) with architecture [2, 32, 32, 1] nodes/layer.
            For the optimization, we used the ADAM optimizer with an initial learning rate of $10^{-3}$ modulated by an exponential decay with multiplicative factor $\gamma=0.99999$. The training was performed for 5000 epochs in the first iteration and for $\sim$20000 epochs for the others. 
            The $\alpha$ hyperparameter in the loss function was set to 0.1.
            The number of iterations, the corresponding dataset size, and the $\lambda$ and the OPES \texttt{BARRIER} used in the biased simulations are summarized in Table~\ref{sup_tab:muller_iterations} alongside the lowest value obtained for functional $K_m$ (e.q., the variational loss term $L_v$), which provides a quality and convergence measure, the simulation time $t_s$ and the output sampling time $t_o$.

            To have a direct comparison with the reference numerical result ${K}_m=4.18$, which is taken from Ref.~\citenum{kang2024computing}, the reported ${K}_m$ values are computed on the same ideal dataset described in Ref.~\citenum{kang2024computing}, which was obtained from a homogeneous grid (that is, 200*200 evenly distributed points) in the relevant part of the Cartesian space (that is,  -1.4<x<1.1 and -0.25<y<2.0 ) assigned with their analytical Boltzmann weights.
            
            \begin {table}[h!]
                \caption {\textbf{Summary M\"uller-Brown} Summary of the iterative procedure for M\"uller-Brown potential.} 
                \label{sup_tab:muller_iterations}
                \vspace{-5mm}
                \begin{center}
                \begin{tabular}{ |c|c|c|c|c|c|c| } 
                 \hline
                 Iteration & Dataset size & $K_m$ [au] & OPES \texttt{BARRIER} [$k_BT$] & $\lambda$ & $t_s$ [au] & $t_o$ [au] \\ 
                 \hline
                    0   & 4000 & 127.2   & - & - & 2*400000  & 200 \\
                    1   & 22000 & 4.25  & 20 & 1 & 2*5000000 & 500 \\
                    2   & 40000 & 4.24  & 20 & 1 & 2*5000000 & 500 \\  
                 \hline
                \end{tabular}
                \end{center}
            \end {table}

    \subsection{Additional figures}
\vspace{-5mm}
    \begin{figure}[h!]
        \centering
        \includegraphics[width=0.8\linewidth]{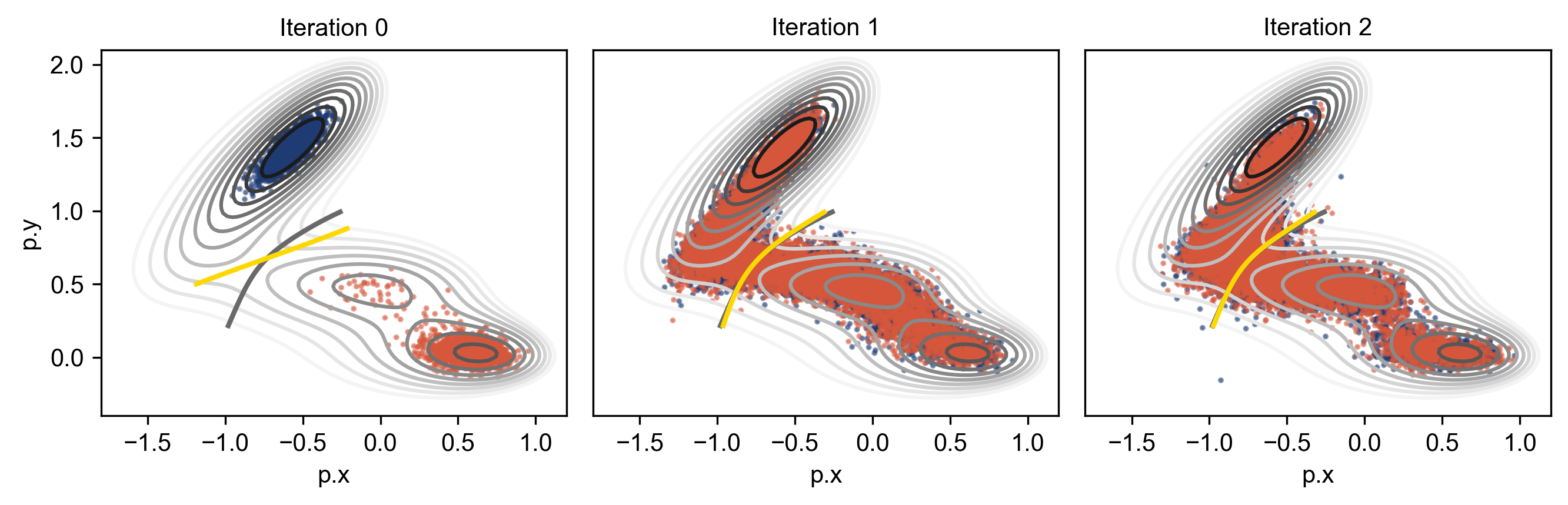}
        \caption{Sampling at different iterations on the procedure on the M\"uller-Brown potential.
        Points obtained starting the simulation in state A are depicted in blue, and those sampled starting from state B are in red.
        For the relevant part of the phase space, the 0.5 isolines of a numerical estimate of the committor and the committor estimate at that iteration are provided as a yellow and grey line, respectively.}
        \label{sup_fig:muller_sampling}
    \end{figure}

\clearpage
\section{Alanine Dipeptide - Additional information}
    \subsection{Computational details}
        \paragraphtitle{Simulations details}
            The alanine dipeptide (Ace-Ala-Nme) simulations in vacuum have been done using the GROMACS-2021.5~\cite{abraham2015gromacs} MD engine patched with PLUMED~\cite{tribello2014plumed, plumed2019promoting} and the Amber99-SB~\cite{amber2013} force field with a 2 fs timestep. 
            The Langevin dynamics is sampled with damping coefficient $\gamma_i= \frac {m_i}{\tau-t}$ with $\tau-t = 0.05$ ps and target temperature 300 K.

    \paragraphtitle{Committor model training details}
        To model the committor function $q_\theta(\textbf{x})$ at each iteration, we used the 45 distances between all heavy atoms as inputs of a neural network (NN) with architecture  [45, 32, 32, 1] nodes/layer. 
        For the optimization, we used the ADAM optimizer with an initial learning rate of $10^{-3}$ modulated by an exponential decay with multiplicative factor $\gamma=0.9999$. 
        The training was performed for 5000 epochs in the first iteration and for $\sim$40000 epochs for the others. 
        The $\alpha$ hyperparameter in the loss function was set to 10. 
        The number of iterations, the corresponding dataset size, and the $\lambda$ and the OPES \texttt{BARRIER} used in the biased simulations are summarized in Table~\ref{sup_tab:alanine_iterations} alongside the lowest value obtained for the functional $K_m$, which provides a quality and convergence measure, the simulation time $t_s$ and the output sampling time $t_o$.
            \begin {table}[h!]
                \caption {\textbf{Summary Alanine} Summary of the iterative procedure for Alanine.} \label{sup_tab:alanine_iterations}
                \begin{center}
                \begin{tabular}{ |c|c|c|c|c|c|c| } 
                 \hline
                 Iteration & Dataset size & $K_m$ [au] & OPES \texttt{BARRIER} [kJ/mol] & $\lambda$ & $t_s$ [ns] & $t_o$ [ps] \\ 
                 \hline
                    0   & 4000 & 5614.68 & - & -   & 4 & 0.4 \\
                    1   & 22000 & 6.24  & 30 & 0.8 & 2*10 & 1 \\
                    2   & 22000 & 1.91 & 30 & 0.8 & 2*10 & 1 \\ 
                    3   & 22000 & 1.08 & 30 & 0.8 & 2*10 & 1 \\  
                 \hline
                \end{tabular}
                \end{center}
            \end {table}

    \subsection{Additional figures}

    \begin{figure}[h!]
        \centering
        \includegraphics[width=1\linewidth]{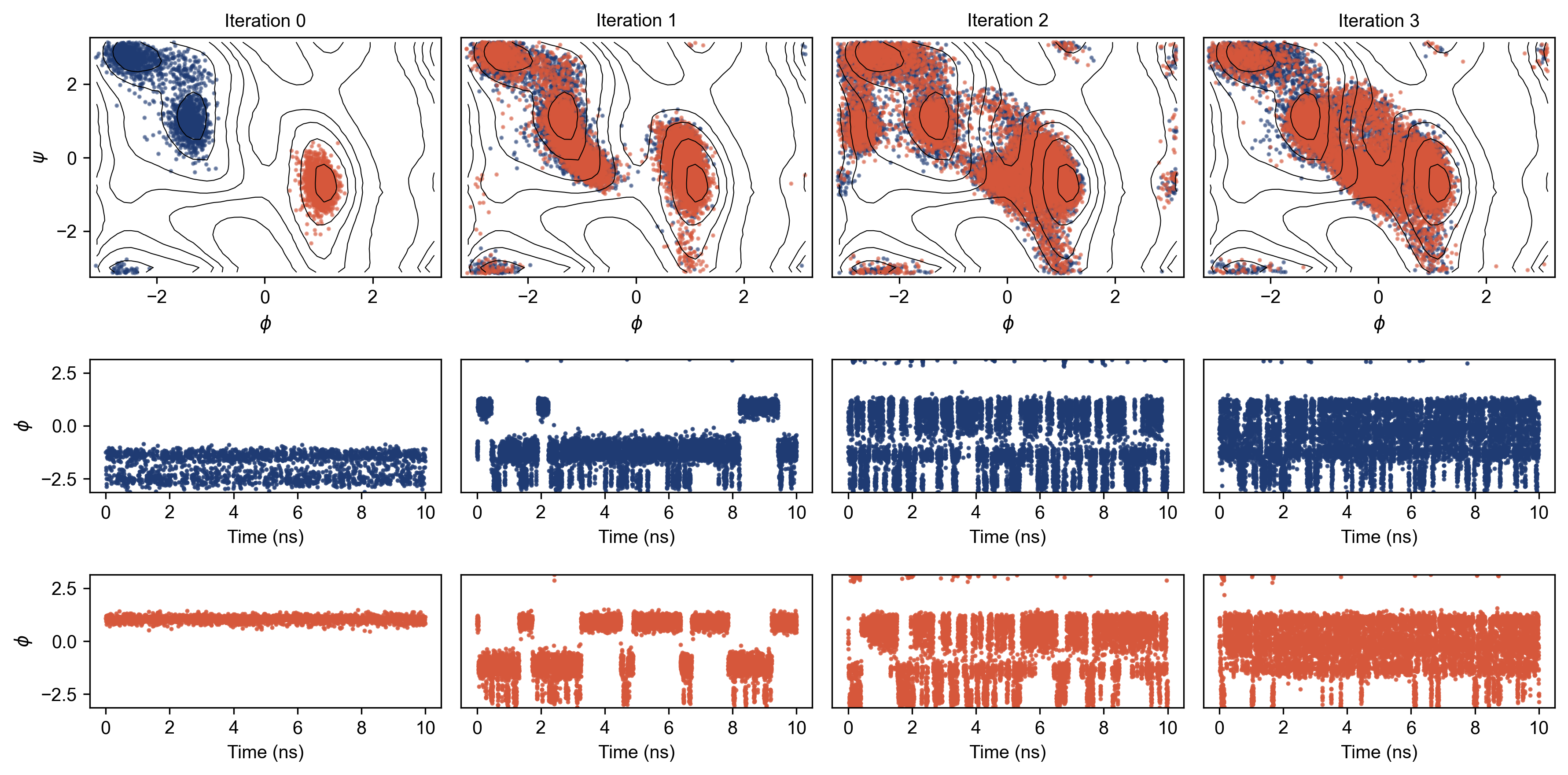}
        \caption{\textbf{Iterations sampling Alanine} Sampling at different iterations on the procedure on Alanine.
        Top row: Scatter plot of the sampled points in the $\phi\psi$ space. Points obtained starting the simulation in state A (left) are depicted in blue, and those sampled starting from state B (right) are in red..
        Middle row: Time series of the $\phi$ value from a trajectory started from state A.
        Bottom row: Time series of the $\phi$ value from a trajectory started from state B.}
        \label{sup_fig:alanine_sampling}
    \end{figure}

\clearpage
\section{Double path potential - Additional information}
    \subsection{Computational details}
        \paragraphtitle{Simulations details}
            The double path potential energy surface, $U(x, y)$, is defined as a function of the Cartesian coordinates $x$ and $y$ 
                \begin{equation}
                \begin{split}
                     U(x,y) = 10\bigg(2+\frac{4x^4}{3}-2y^2+y^4+\frac{10x^2(y^2-1)}{3}\bigg) \\
                     + 7\exp \bigg( -\frac{(x+0.7)^2 + (y-0.8)^2}{0.4^2}\bigg) \\
                     + \exp \bigg( -\frac{(x-1)^2 + (y+0.3)^2}{0.4^2}\bigg) \\ 
                     - 6\exp\bigg( -\frac{(x+1)^2 + (y+0.6)^2)}{0.4^2}\bigg) \\
                     -2.35906
                    \label{seq:double_path_formula}
                \end{split}
                \end{equation}
            
            The simulations of the diffusion of an ideal particle of mass 1 have been performed using Langevin dynamics based on the Bussi-Parrinello algorithm~\cite{bussi2007accurate} as implemented in the \texttt{ves\_md\_linearexpansion}~\cite{valsson2014variational} module of PLUMED.~\cite{tribello2014plumed, plumed2019promoting}
            The damping constant in the Langevin equation was set to  10/time-unit. 
            The time unit was defined arbitrarily and corresponds to 200 timesteps and natural units ($k_BT = 1$) were used in all the calculations.

        \paragraphtitle{Committor model training details}
            To model the committor function $q_\theta(\textbf{x})$ at each iteration, we used the x and y Cartesian coordinates of the diffusing particle as inputs of a neural network (NN) with architecture [2, 32, 32, 1] nodes/layer.
            For the optimization, we used the ADAM optimizer with an initial learning rate of $10^{-3}$ modulated by an exponential decay with multiplicative factor $\gamma=0.99999$. The training was performed for 5000 epochs in the first iteration and for $\sim$20000 epochs for the others. 
            The $\alpha$ hyperparameter in the loss function was set to 10.
            The number of iterations, the corresponding dataset size, and the $\lambda$ and the OPES \texttt{BARRIER} used in the biased simulations are summarized in Table~\ref{sup_tab:double_path_iterations} alongside the lowest value obtained for functional $K_m$ (e.q., the variational loss term $L_v$), which provides a quality and convergence measure, the simulation time $t_s$ and the output sampling time $t_o$. 
            \begin {table}[h!]
                \caption {\textbf{Summary double-path potential} Summary of the iterative procedure for double path potential.} \label{sup_tab:double_path_iterations}
                \vspace{-5mm}
                \begin{center}
                \begin{tabular}{ |c|c|c|c|c|c|c| } 
                 \hline
                 Iteration & Dataset size & $K_m$ [au] & OPES \texttt{BARRIER} [$k_BT$] & $\lambda$ & $t_s$ [au] & $t_o$ [au] \\ 
                 \hline
                    0   & 4000 &  35.77  & - & - & 2*5000000 & 500 \\
                    1   & 22000 &  1.38 & 20 & 1 & 2*5000000 & 500\\
                    2   & 40000 & 1.37  & 20 & 1 & 2*5000000 & 500 \\  
                 \hline
                \end{tabular}
                \end{center}
            \end {table}
    
    \vspace{-5mm}
    \subsection{Additional figures}
    \vspace{-5mm}
        \begin{figure}[h!]
            \centering
            \includegraphics[width=0.8\linewidth]{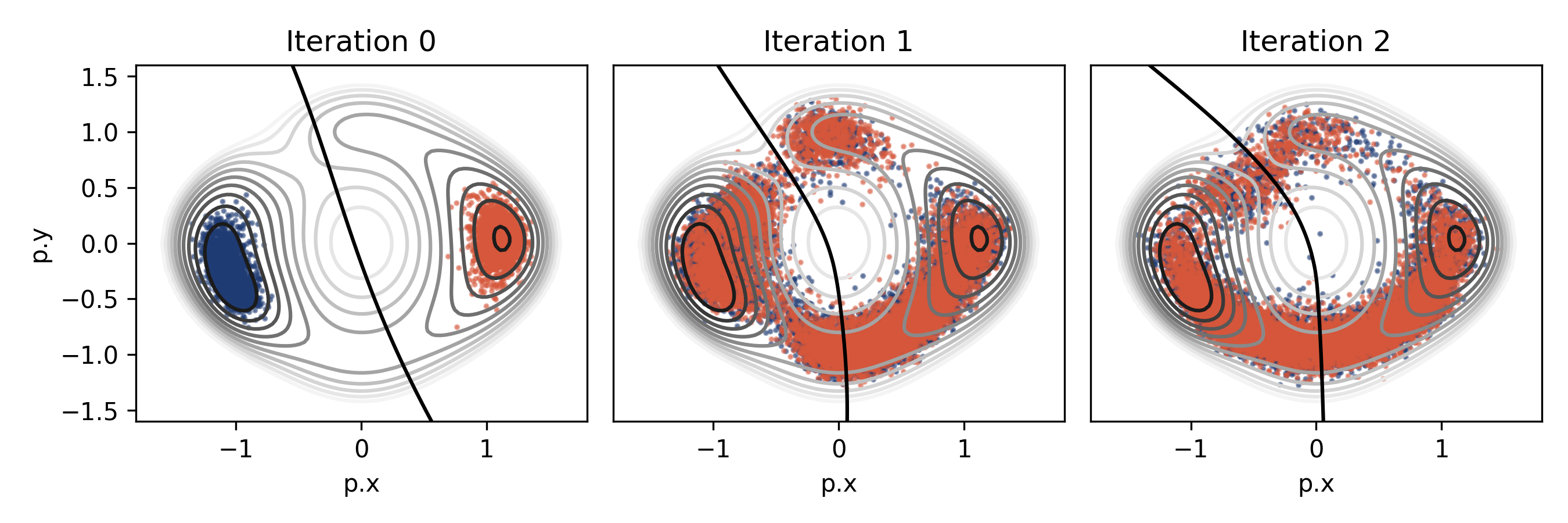}
            \caption{\textbf{Iterations sampling double-path} Sampling at different iterations of the procedure on the double path potential.
            Points obtained starting the simulation in state A (left) are depicted in blue, and those sampled starting from state B (right) are in red.
            The 0.5 value of the committor estimate at that iteration is provided as a black line.}
            \label{sup_fig:double_path_sampling}
        \end{figure}

\clearpage
\section{Chignolin - Additional information}
    \subsection{Computational details}
        \paragraphtitle{Simulations details}
            For the study of folding and unfolding of chignolin (CLN025 peptide sequence Tyr-Tyr-Asp-Pro-Glu-Thr-Gly-Thr-Trp-Tyr) in explicit solvent, we performed our simulations using GROMACS v2021.5~\cite{abraham2015gromacs} patched with PLUMED,~\cite{tribello2014plumed, plumed2019promoting} the CHARMM22$^*$~\cite{piana2011robust} force field, and the solvent has been modeled by the CHARMM TIP3P~\cite{mackerell1998tip} force field, sharing the same setup used for long unbiased simulations on this system~\cite{lindorff2011fast} to have a direct comparison with those results. 
            For the same reason, we kept the simulation condition consistent with that work.
            All simulations were performed with an integration time step of 2 fs and sampling NVT ensemble at 340K. Asp, Glu residues, as well as the N- and C-terminal amino acids are simulated in their charged states. The simulation box contains 1,907 water molecules, together with two sodium ions that neutralize the system. The linear constraint solver algorithm is applied to every bond involving H atoms, and electrostatic interactions are computed via the particle mesh Ewald scheme, with a cutoff of 1 nm for all nonbonded interactions.

        \paragraphtitle{Committor model training details}
            To model the committor function $q_\theta(\textbf{x})$ at each iteration, we used the 210 distances suggested by Ref.~\citenum{bonati2021deep} as inputs of a neural network (NN) with architecture  [210, 64, 32, 1] nodes/layer.
            For the optimization, we used the ADAM optimizer with an initial learning rate of $10^{-3}$ modulated by an exponential decay with multiplicative factor $\gamma=0.99995$. 
            The training was performed for $\sim$30000 epochs. The $\alpha$ hyperparameter in the loss function was set to 1.
            The number of iterations, the corresponding dataset size, and the $\lambda$ and the OPES \texttt{BARRIER} used in the biased simulations are summarized in Table~\ref{sup_tab:chignolin_iterations} alongside the lowest value obtained for the functional $K_m$, which provides a quality and convergence measure, the simulation time $t_s$ and the output sampling time $t_o$.
            \begin {table}[h!]
                \caption {\textbf{Summary Chignolin} Summary of the iterative procedure for chignolin.} \label{sup_tab:chignolin_iterations}
                \begin{center}
                \begin{tabular}{ |c|c|c|c|c|c|c| } 
                 \hline
                 Iteration & Dataset size & $K_m$ [au] & OPES \texttt{BARRIER} [kJ/mol] & $\lambda$ & $t_s$ [ns] & $t_o$ [ps] \\ 
                 \hline
                    0   & 20000 &  27.36  & -& -     & 2*100 & 10\\
                    1   & 220000 &  4.43 & 20& 0.3  & 4*250 & 5\\
                    2   & 220000 &  3.64 & 20& 0.3  & 4*250 & 5\\
                    3   & 220000 &  3.79 & 20& 0.3  & 4*250 & 5\\
                    4   & 220000 &  2.32 & 20& 0.3  & 4*250 & 5\\
                    5   & 220000 &  2.17 & 20& 0.3  & 4*250 & 5\\   
                 \hline
                \end{tabular}
                \end{center}
            \end {table}

    \subsection{Additional results}
    \paragraphtitle{Comparison between different descriptor sets}
        In Ref.~\citenum{kang2024computing}, we trained the committor model using as inputs the 45 distances between $\alpha$-C atoms, whereas here, we used the same set of 210 distances identified in Ref.~\citenum{bonati2021deep}.
        To compare the two descriptor sets, we optimized the committor model on the same dataset using the two sets, obtaining a $K_m$ value of 2.17 for the 210 distances and 18.5 for the 45 distances.
        This result clearly shows how the greater variational flexibility introduced by side-chain information allows for capturing more features of the folding process.

    \paragraphtitle{LASSO Analysis}
        To analyze the relevance of the input feature in the trained model, we applied the LASSO regression analysis~\cite{novelli2022lasso} on several bins along $z$ values. 
        The results of this analysis are summarized in Tab.~\ref{sup_tab:chignolin_lasso}, where the 5 most important descriptors for each bin are reported alongside the corresponding bin range.
        
            \begin {table}[h!]
                \caption {\textbf{LASSO summary Chignolin} Summary of LASSO analysis for chignolin. The atoms involved in the distances used as descriptors are indicated according to the indexes used in PLUMED (upper table) and their names from the PDB (lower table).} \label{sup_tab:chignolin_lasso}
                \begin{center}
                \begin{tabular}{ |c|c| } 
                 \hline
                 z range & Top 5 ranking distances \\ 
                 \hline
                    $-5.5\sim -4.3$  & $d_{102-139},d_{107-139},d_{102-141},d_{109-131},d_{78-141}$\\
                    $-4.3\sim -3.2$ & $d_{102-139},d_{102-141},d_{107-139},d_{78-141},d_{109-131}$\\
                    $-3.2\sim -2.0$  &$d_{36-102},d_{23-145},d_{36-70},d_{7-145},d_{45-147}$\\
                    $-2.0\sim -0.8$  &$d_{22-164},d_{37-69},d_{23-164},d_{37-109},d_{43-120}$\\
                    $-0.8\sim 0.3$ &   $d_{22-164},d_{23-164},d_{36-105},d_{23-145},d_{37-69}$\\
                    $0.3\sim 1.5$  &  $d_{23-145},d_{43-120},d_{23-106},d_{36-144},d_{44-152}$\\
                    $1.5\sim 2.7$ &$d_{43-120},d_{44-152},d_{23-145},d_{37-109},d_{37-66}$\\
                    $2.7\sim 3.8$&  $d_{52-86},d_{63-105},d_{31-100},d_{41-141},d_{56-105}$\\
                    $3.8\sim 5.0$ &   $d_{43-120},d_{31-115},d_{52-86},d_{49-111},d_{41-144}$\\
                    $5.0\sim 6.1$& $d_{34-128},d_{52-94},d_{52-83},d_{1-152},d_{11-165}$\\
                 \hline
                \end{tabular}
                
                \vspace{8mm}
                
                \begin{tabular}{ |c|c| } 
                 \hline
                 z range & Top 5 ranking distances \\ 
                 \hline
                    $-5.5\sim -4.3$& Gly7-CA:Trp9-CZ2, Thr8-N:Trp9-CZ2,  Gly7-CA:Trp9-CH2, Thr8-CA:Trp9-NE1,  Glu5-CG:Trp9-CH2\\
                    $-4.3\sim -3.2$& Gly7-CA:Trp9-CZ2, Gly7-CA:Trp9-CH2, Thr8-N:Trp9-CZ2,  Glu5-CG:Trp9-CH2,  Thr8-CA:Trp9-NE1\\
                    $-3.2\sim -2.0$& Tyr2-CZ:Gly7-CA,  Tyr1-O:Tyr10-N,   Tyr2-CZ:Pro4-O,   Tyr1-CB:Tyr10-N,   Asp3-N:Tyr10-CA\\
                    $-2.0\sim -0.8$& Tyr1-C:Tyr10-C,   Tyr2-OH:Pro4-C,   Tyr1-O:Tyr10-C,   Tyr2-OH:Thr8-CA,   Tyr2-C:Thr8-O\\
                    $-0.8\sim 0.3$ & Tyr1-C:Tyr10-C,   Tyr1-O:Tyr10-C,   Tyr2-CZ:Gly7-C,   Tyr1-O:Tyr10-N,    Tyr2-OH:Pro4-C\\
                    $0.3\sim 1.5$  & Tyr1-O:Tyr10-N,   Tyr2-C:Thr8-O,    Tyr1-O:Gly7-O,    Tyr2-CZ:Trp9-O,    Tyr2-O:Tyr10-CG\\
                    $1.5\sim 2.7$  & Tyr2-C:Thr8-O,    Tyr2-O:Tyr10-CG,  Tyr1-O:Tyr10-N,   Tyr2-OH:Thr8-CA,   Tyr2-OH:Pro4-CG\\
                    $2.7\sim 3.8$  & Asp3-CG:Thr6-N,   Pro4-CB:Gly7-C,   Tyr2-CG:Gly7-N,   Tyr2-CE2:Trp9-CH2, Asp3-O:Gly7-C\\
                    $3.8\sim 5.0$  & Tyr2-C:Thr8-O,    Tyr2-CG:Thr8-CG2, Asp3-CG:Thr6-N,   Asp3-CB:Thr8-CB,   Tyr2-CE2:Trp9-O\\
                    $5.0\sim 6.1$  & Tyr2-CE1:Trp9-CG, Asp3-CG:Thr6-CG2, Asp3-CG:Glu5-OE2, Tyr1-N:Tyr10-CG,   Tyr1-CD1:Tyr10-HE2\\
                 \hline
                \end{tabular}
                \end{center}
            \end {table}

\newpage
    \paragraphtitle{Free energy surfaces}
        \begin{figure}[h!]
            \centering
            \includegraphics[width=0.8\linewidth]{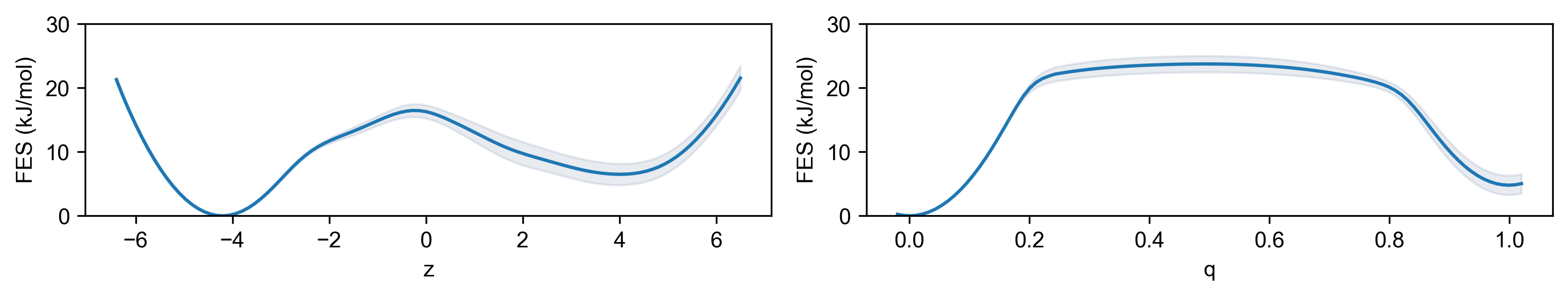}
            \caption{\textbf{Free energy surface Chignolin} Free energy estimates for Chignolin projected along $z$ and $q$ obtained averaging on 6 independent simulations: 3 started from the folded state and 3 from the unfolded one. 
            The average value and the relative standard deviation are reported as a solid line and a shaded region, respectively.}
            \label{sup_fig:chignolin_fes_z_q}
        \end{figure}

\clearpage
\section{Calixarene - Additional information}
    \subsection{Computational details}
        \paragraphtitle{Simulations details}
            We performed the simulations with GROMACS~\cite{abraham2015gromacs} v2021.5 in combination with the PLUMED~\cite{tribello2014plumed, plumed2019promoting} plugin. We use the GAFF~\cite{wang2004development} force field with RESP~\cite{bayly1993well} charges and the TIP3P~\cite{jorgensen1983comparison} water model. 
            We set the timestep to 2 fs and the temperature at 300 K via a velocity rescale thermostat~\cite{bussi2007velocity} with a time constant of 0.1 ps.
            The simulation box is cubic with a side of 40.27 $\si{\angstrom}$ and it contains 2100 water molecules in solution together with the host OAMe and the chosen guest molecule. 
            Sodium ions are included to counterbalance excess charges. 
            At every simulation step, the coordinates are aligned so that the vertical axis of the box coincides with the binding axis $h$, and the simulation box is centered on the virtual atom V1.

        \paragraphtitle{The funnel restraint}
        \label{sup_sec:funnel_restraint}
            In our simulations, we used a funnel restraint~\cite{limongelli2013funnel} equivalent to the one previously employed by Refs.\citenum{rizzi2021role,bhakat2017resolving,perez2019local} on the same system.
            Here, we summarize the details of such a restraint, while more details can be found in the original works.
            The funnel limits the space available to the ligand in the unbound state U by confining it to a cylindrical volume above the binding site. 
            As the ligand approaches the binding site, the funnel restraint becomes wider so that its presence does not affect the binding process itself. 
            Having aligned with PLUMED the system to a reference configuration where the binding axis is found along the vertical axis, we define $h$ as the projection on the binding axis of the center of the carbon atoms of each ligand and $r$ as its radial component.
            When $h>10 \si{\angstrom}$, the funnel surface is a cylinder with radius $R_{cyl} = 2 \si{\angstrom}$ with its axis along the vertical direction. 
            When $h<10 \si{\angstrom}$, the funnel opens into an umbrella-like shape with a 45 degree angle whose surface is defined by $r=12-h$.
            The force that, for a displacement x, pushes the ligand away from the funnel’s surface is harmonic $-k_Fx$ with $kF=20 $ kJ/mol $\si{\angstrom}^{-2}$. 
            A further harmonic restraint is applied on $h$  to prevent the ligand from getting too far from the host, reaching the upper boundary of the simulation box. 
            The corresponding force is $-k_U(h-18)$ for $h > 18\si{\angstrom}$ and $k_U=40 $ kJ/mol $\si{\angstrom}^{-2}$.

            During training, we set boundaries further to state U so that the labeled configurations used to impose the boundary conditions will not influence the committor training in the subsequent iterative simulations.
            We apply the funnel restraint described above and two additional harmonic restraints $-k_U(h-20)$ for $h > 20 \si{\angstrom}$ and $-k_U(h-18)$ for $h < 18 \si{\angstrom}$, with $k_U = 20$ kJ/mol.
        
            Because of the funnel presence, the free energy difference between the bound and the true unbound state that we extract from enhanced sampling simulations needs a correction that can be calculated as:
                \begin{equation}
                    \Delta G = -\frac{1}{\beta} \log \left( C_0 \pi R_{\text{cyl}}^2 \int_B dh \exp\left[-\beta \left(W(h) - W_U\right)\right] \right)
                \end{equation}
            where $\beta$ = 1/($k_B$T), $C_0$ = 1/1660 $\si{\angstrom}^{-3}$ is the standard concentration, $h$ is the coordinate along the funnel’s axis, $W(h)$ is the free energy along the funnel axis and $W_U$ its reference value in state U. 
            More precisely, we define $W_U$ as the average free energy value in the interval 1.6  $\si{\angstrom}$ < $h$ < 1.8  $\si{\angstrom}$.
            The integral is computed over the state B region that we define as 0.3  $\si{\angstrom}$ < $h$ < 0.8  $\si{\angstrom}$.

        \paragraphtitle{Committor model training details}
            To model the committor function $q_\theta(\textbf{x})$ at each iteration, we used the same water coordination numbers used in Ref.~\citenum{rizzi2021role}, 8 with respect to 2.5\AA-spaced virtual atoms (V) along the binding axis and 4 with respect to 4 atoms on the ligand molecule (M) (2 on the ring (1 and 2), 2 on the terminal atoms (3 and 4)), and a set of distances between the same 4 ligand atoms and the binding pocket, instead of the $h$ projection only, as inputs of a neural network (NN) with architecture [16, 32, 32, 1] nodes/layer.
            For the optimization, we used the ADAM optimizer with an initial learning rate of $10^{-3}$ modulated by an exponential decay with multiplicative factor $\gamma=0.99995$.
            The training was performed for $\sim$30000 epochs. 
            The $\alpha$ hyperparameter in the loss function was set to 1.
            In the loss function, we assigned to the virtual points along the binding axis used for the computation of water coordination numbers an \emph{infinite} mass, that is, $10^6$.
            The number of iterations, the corresponding dataset size, and the $\lambda$ and the OPES \texttt{BARRIER} used in the biased simulations are summarized in Table~\ref{sup_tab:calixarene_iteration} alongside the lowest value obtained for the functional $K_m$, which provides a quality and convergence measure, the simulation time $t_s$ and the output sampling time $t_o$.

            \begin {table}[h!]
                \caption {\textbf{Summary calixarene} Summary of the iterative procedure for calixarene.} \label{sup_tab:calixarene_iteration}
                \begin{center}
                \begin{tabular}{ |c|c|c|c|c|c|c| } 
                 \hline
                 Iteration & Dataset size & $K_m$ [au] & OPES \texttt{BARRIER} [kJ/mol] & $\lambda$ & $t_s$ [ns] & $t_o$ [ps] \\ 
                 \hline
                    0   & 2000 &  27400  & -& -     & 2*10 & 10\\
                    1   & 22000 &  137 & 50& 0.6  & 2*100 & 10\\
                    2   & 22000 &  333 & 50& 0.6  & 2*100 & 10\\
                    3   & 22000 &  242 & 50& 0.6  & 2*100 & 10\\
                    4   & 22000 &  5.16 & 50& 0.6  & 2*100 & 10\\
                    5   & 22000 &  4.56 & 50& 0.6  & 2*100 & 10\\
                 \hline
                \end{tabular}
                \end{center}
            \end {table}

    \newpage
    \subsection{Additional results}
    \paragraphtitle{Feature relevance analysis}
        \begin{figure}[h!]
            \centering
            \includegraphics[width=0.8\linewidth]{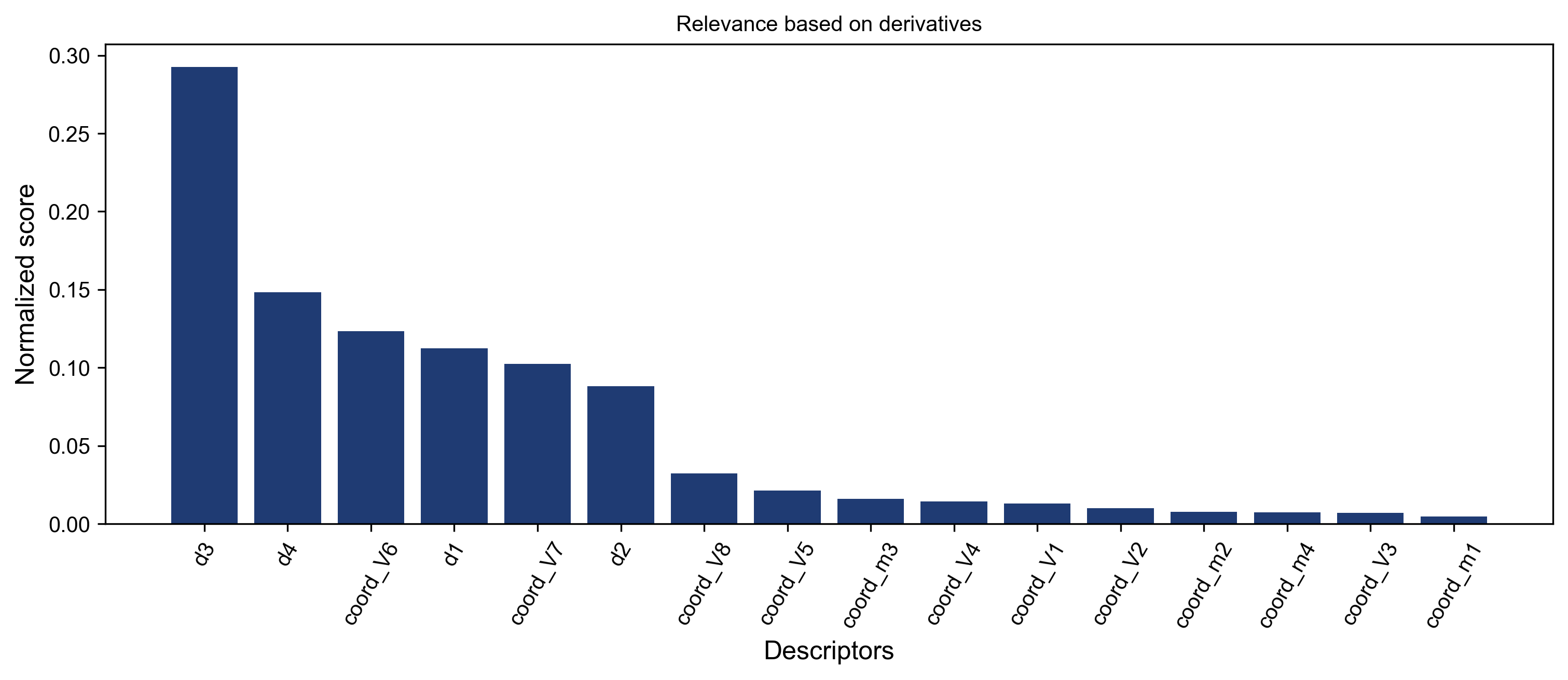}
            \caption{\textbf{Feature relevance calixarene} Descriptors ranking for the committor model of calixarene trained using the descriptors listed in the computational details.}
            \label{sup_fig:calixarene_feature_relevance}
        \end{figure}
        
    \paragraphtitle{Free energy surfaces}
        \begin{figure}[h!]
            \centering
            \includegraphics[width=1\linewidth]{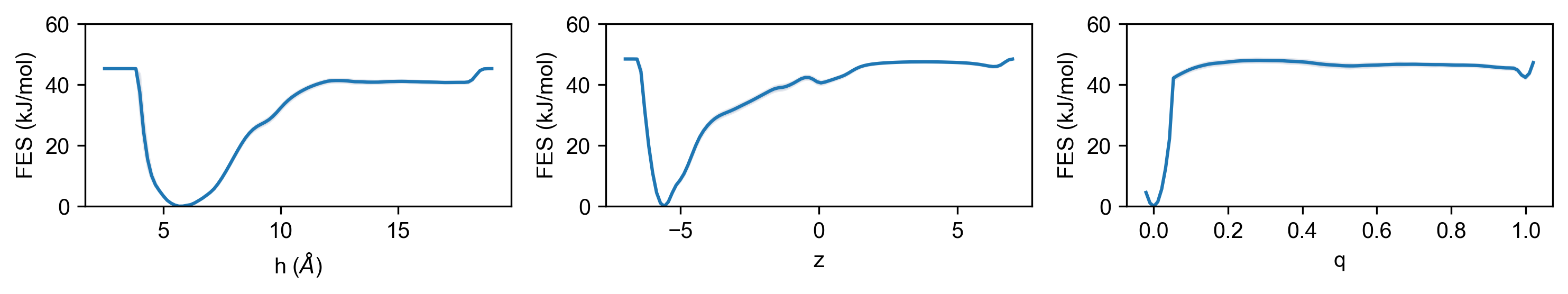}
            \caption{\textbf{1D free energy surface calixarene} Free energy estimates for calixarene projected along the projection of the ligand molecule on the binding axis $h$, $z$ and $q$.}
            \label{sup_fig:Calixarene_fes_z_q}
        \end{figure}

        \begin{figure}[h!]
            \centering
            \includegraphics[width=0.5\linewidth]{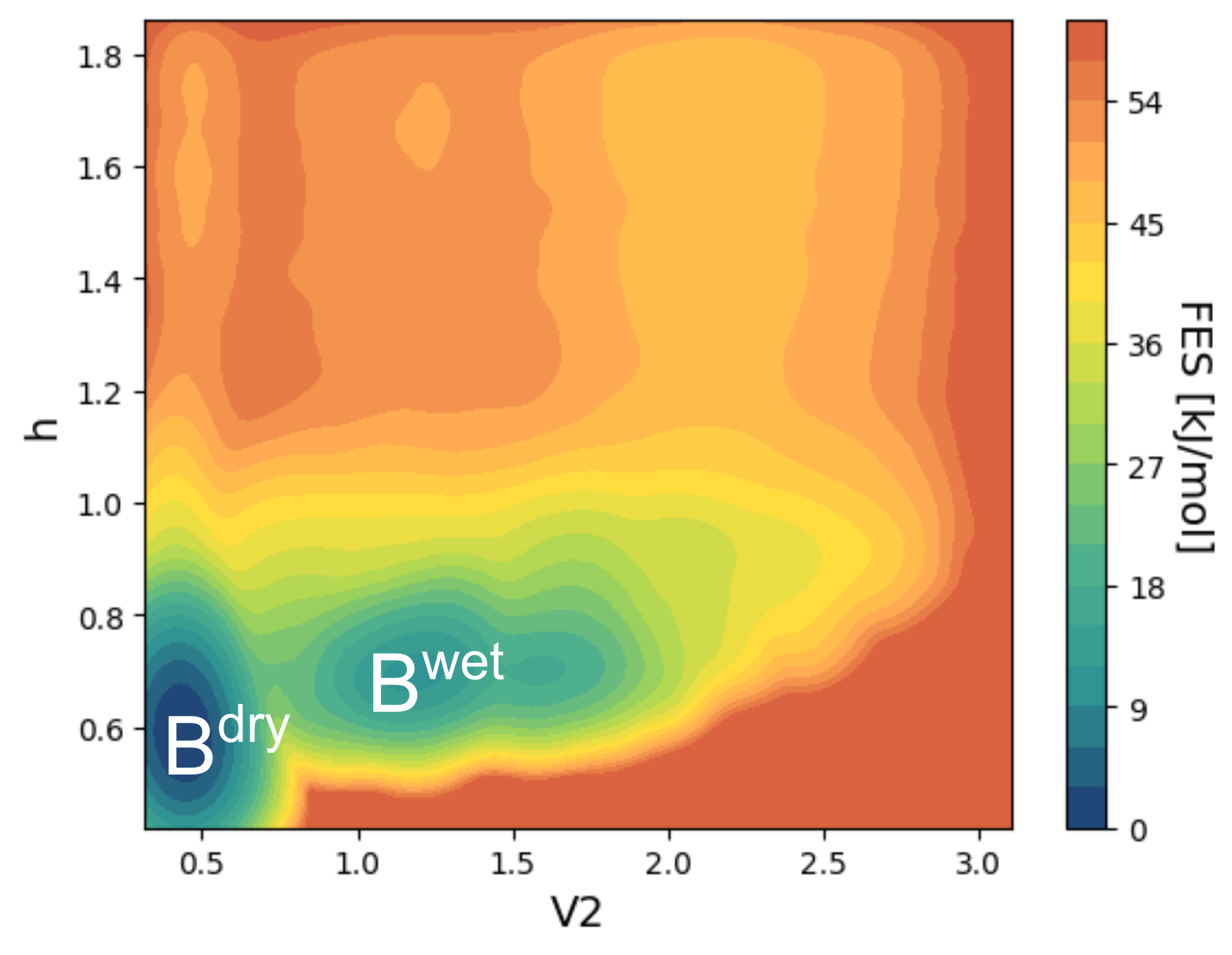}
            \caption{\textbf{2D free energy surface calixarene} Two-dimensional free energy surface for calixarene projected in the plane defined by the water coordination number of a virtual point inside the binding cavity (V2) and the projection of the ligand molecule on the binding axis (h). }
            \label{sup_fig:Calixarene_2DFES}
        \end{figure}

    \newpage
    \paragraphtitle{Kolmogorov Ensemble}

        \begin{figure}[h!]
            \centering
            \includegraphics[width=0.5\linewidth]{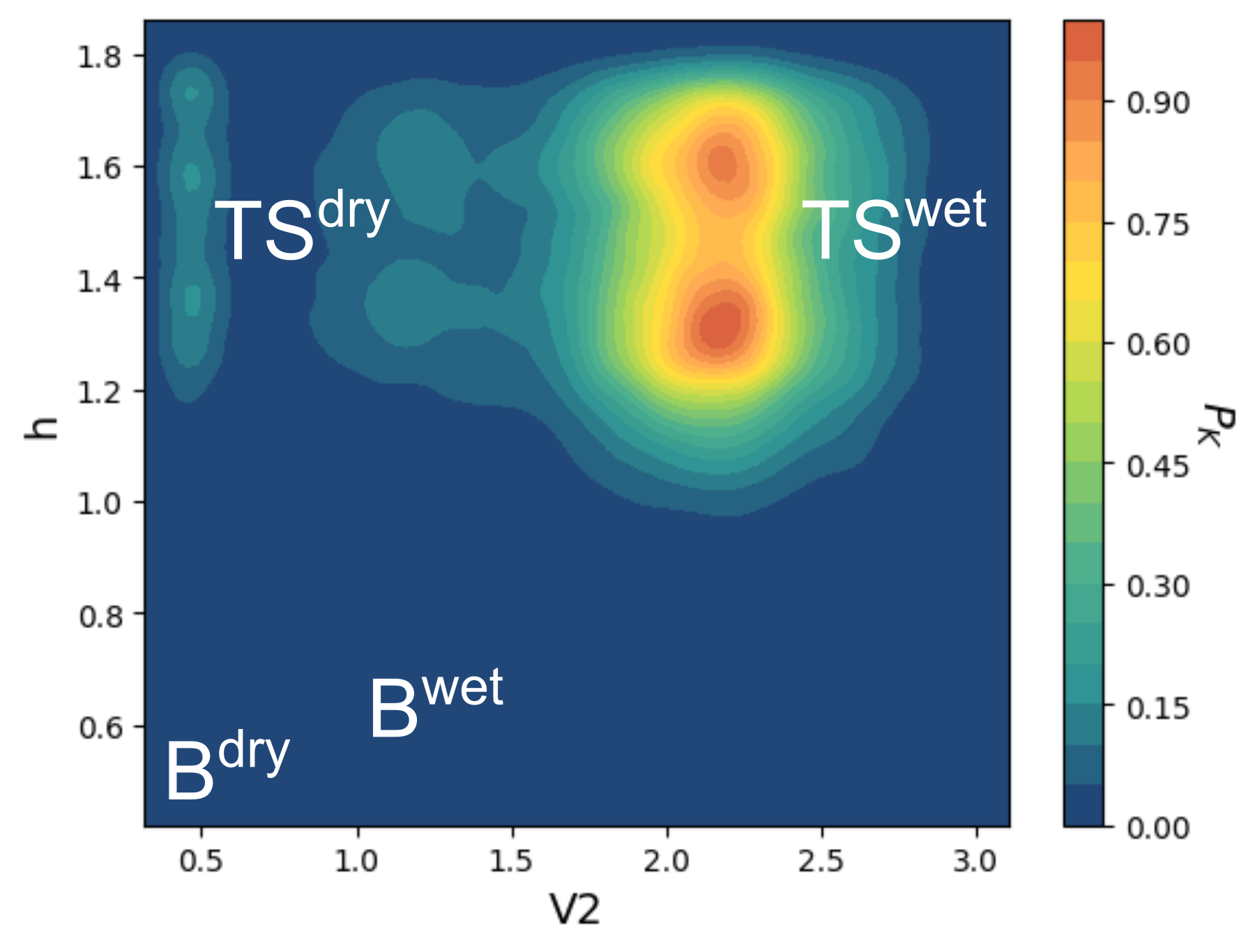}
            \caption{\textbf{Kolomogorov distribution calixarene} Two-dimensional Kolmogorov probability $p_K$ projection for calixarene projected in the plane defined by the water coordination number of a virtual point inside the binding cavity (V2) and the projection of the ligand molecule on the binding axis (h). }
            \label{sup_fig:Calixarene_2D_K}
        \end{figure}



\end{document}